\documentclass[pra,twocolumn]{revtex4}
\usepackage{latexsym,amssymb,amsmath,graphicx}
\usepackage{color}

\definecolor{ascolor}{RGB}{45, 119, 43}

\begin{document}

\title{Design of the coherent diffusive photon gun for generating non-classical states}
\author{{M. Thornton}$^1$, {A. Sakovich}$^2$, A. Mikhalychev$^2$, J. D. Ferrer$^1$, P. de la Hoz$^1$, {N. Korolkova}$^1$, D. Mogilevtsev$^2$}
\affiliation{$^1$ School of Physics and Astronomy, University of St Andrews,
North Haugh, St Andrews KY16 9SS, UK, \\
$^2$Institute of Physics, Belarus National Academy of Sciences, Nezavisimosti Ave. 68-2, Minsk 220072 Belarus}
\begin{abstract}
We suggest and discuss a concept of deterministic integrated source of non-classical light based on the coherent diffusive photonics, a coherent light flow in a system of dissipatively coupled waveguides. We show how this practical quantum device can be realized with a system of single-mode waveguides laser-inscribed in nonlinear glass. We describe a hierarchy of models, from the complete multi-mode model of the waveguide network to the single mode coupled to a bath, analyze  the conditions for validity of the simplest single-mode model and demonstrate feasibility  of the generation of bright sub-Poissonian light states merely from  a coherent input.  Notably, the generation of  non-classical states occurs at the initial stages of the dynamics, and can be accounted for in the linear model that allows us to circumvent the prohibiting computational complexity of the exact full quantum representation.
\end{abstract}
%\nopacs
\maketitle

\section{Introduction}

Engineered loss has already turned into a powerful and intensively researched tool for quantum state manipulation. With the help of engineered reservoirs, it is possible to drive a target system into a desired state, to generate and protect entanglement, to implement computation, and to transfer quantum states \cite{carmichael,cirac,metel}. Engineered reservoirs have become a popular and fruitful research direction for QED \cite{clark}, cold atoms  and trapped ions \cite{zoller,blatt}, Rydberg atoms \cite{carr,rog}, and for super-conducting circuits \cite{science2015,kimchi,kapit} due to the large spectra of possibilities to design systems with necessary coupling between components and strong effective nonlinearities. For example, by devising strong self-Kerr and cross-Kerr interactions, and combining them with two-photon conversion and classical driving, it is possible to produce two-photon loss, and {in this way} create the Schr{\"o}dinger-cat states and safeguard {them against} linear loss  \cite{science2015}.

\begin{figure}[h]
\includegraphics[width=0.9\linewidth]{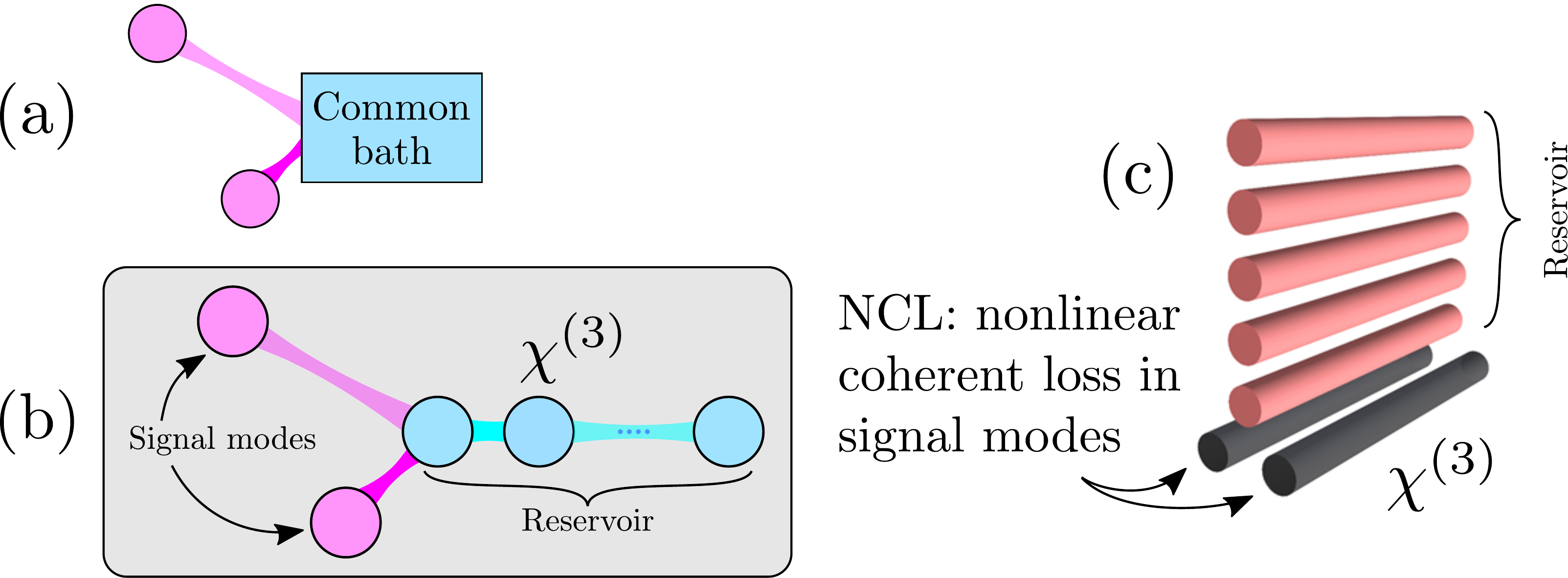}
\caption{{Deterministic photon gun (PhoG) based on the coherent diffusive photonics (CDP) network of single-mode waveguides written in a bulk of nonlinear glass.   (a) Basic scheme behind the PhoG device: two signal modes asymetrically coupled to a common bath. Both modes undergo nonlinear coherent loss (NCL). (b) Implementation of the basic scheme using network of nonlinear waveguides. The common bath, the reservoir, is implemented as a linear array of further waveguides (``tail''). Signal modes (grey in 3D picture (c)) are dissipatively coupled via this common reservoir. The ``tail'' waveguides (pink in 3D picture (c)) interact via conventional evanescent coupling. }}
%(a) The PhoG device: coherent input light is coupled into ``signal" waveguides of the waveguide network laser-inscribed in a few-centimeter long thin glass plate with Kerr $\chi^{\left(3\right)}$ nonlinearity. Depending on the regime chosen, either sub-Poissonian light is observed at the output in a collective mode or entangled photon pairs are produced in two output modes.
\label{fig:1}
\end{figure}
It is considerably harder to engineer losses {in} photonic circuits. Optical nonlinearities are usually quite small and accompanied by rather strong conventional linear loss. However, even in this case there are possibilities to exploit effects of nonlocal loss and devise photonic circuits with quite unusual and potentially very useful features, even in passive optical structures. These possibilities are opened {up} by a new level of precision and controllability reached in systems of coupled waveguides. For example, localization of the light in the small part of a perfect Lieb lattice and diffraction-less propagation was experimentally demonstrated in a system of single-mode waveguides written in a bulk glass \cite{seba1,vicencio}.

These new experimental advances gave rise to recently introduced concept of ``coherent diffusive photonics" (CDP): light-processing {in a specifically designed system of single-mode waveguides  coupled by common loss reservoirs, as realized in practice in \cite{natcom}.} Even with linear glass, dissipative coupling {enables}, for example, light equalization in a waveguide chain{: any input state tends to a completely symmetrized state} over all the modes. In contrast to the common unitary coupling schemes, CDP schemes are quite robust with respect to variations in the coupling length and strength. Further, CDP allows one to  realize an optical router directing light in different arms of the structure by selective excitation of the central control nodes, and CDP can even allow light to be localized in a perfect lattice of dissipatively coupled modes \cite{natcom}.

In this paper we extend the developed CDP schemes to the nonlinear regime, and suggest a new integrated CDP-based device exhibiting   features of a ``photon gun'' \emph{PhoG}: a deterministic generator of nonclassical states, in particular, sub-Poissonian states (Fig.~\ref{fig:1}).  

It is worth noticing that a  true photon gun is still a challenge (for review see \cite{SPSreview1,SPSreview2,SPSreview3}). Due to high technical overhead, the available single photon sources are not readily implementable, mostly do not have well-defined output mode and many of them cannot simultaneously satisfy the crucial figures of merit: purity, indistinguishability (often traded for brightness), and efficiency.  Only very recently top-end quantum dot-based sources could demonstrate simultaneously single-photon purity of $\ge 99\%$, photon indistinguishability of $95-99\%$ and extraction efficiency of up to $ 65\%$ \cite{qdot}. Other alternative, heralded photon sources based on entangled photon pairs production are probabilistic and deliver high indistinguishability at the cost of low brightness. Therefore in many quantum technology applications, experimentalists revert to attenuated coherent states instead of single photons. This is true for most if not all quantum cryptography implementations and many other, including quantum simulations and implementations of quantum gates. Dim coherent states are `cheap', easy to handle and deterministic, however their Poissonian photon number statistics represents a serious limitation, resulting in unwanted contribution from multi-photon components. Much better performance can be achieved using light with sub-Poissonian photon statistics as a quasi-single photon source. For example, the key rates for decoy-state quantum key distribution (QKD) have been explicitely calculated for different sources, and it was  demonstrated that such quasi-single-photon source can drastically raise the key rate in the decoy-state QKD \cite{wang,wang1}.  

We demonstrate that the photon gun device suggested in this paper is potentially able to act either as a deterministic or, upon a slight variation of the scheme, probabilistic quasi-single photon source. We analyze the possibility of realization of these devices with current technology. Whereas approaching deterministic generation of quasi-single photon- or few-photon strongly sub-Poissoinian states still seems challenging, we show that bright sub-Poissonian photon gun is completely feasible and  can be quite robust with respect to experimental imperfections and noise.  

Bright Sub-Poissonian photon gun (PhoG) suggested in this paper carries important advantage of producing non-classical states from merely coherent input, on demand, and in well-defined temporal and spatial modes. Further, PhoG is a versatile quantum source and can be modified to produce correlated photons and other entangled states at the output. Knowingly, non-classical photon number correlations and sub-Poissonian photon statistics can be exploited for quantum sensing, imaging and other metrology tasks  \cite{bio,imaging}.

The basic idea underlying our scheme is to create a photonic circuit exhibiting nonlinear coherent loss (NCL), which is a specific kind of engineered single-photon loss  \cite{mog2010,mog2011}. The principal circuit, Fig.~\ref{fig:1}(a), comprises two signal modes asymmetrically coupled to a common loss reservoir and undergoing NCL. The required light propagation regime can be engineered in different ways. We suggest a nonlinear waveguide network with a chain waveguide structure working as a common reservoir, Fig.~\ref{fig:1}(b), which allows the NCL to be realized in the superposition of two waveguide modes. We analyze its feasibility under realistic Kerr nonlinearity of nonlinear glass and for realistic waveguide parameters. We show that it is feasible to deterministically create bright sub-Poissonian states of light using a compact (cm-sized) integrated CDP device and, furthermore, reach other useful nonclassical output states. The generation of nonclassical light occurs in the regime when a conventional single-photon loss is still not affecting significantly the process of the state generation.

The outline of the paper is as follows.  Firstly, {in Sec. II, we present} the CDP scheme and describe the hierarchy of models which can be  used to account for the state dynamics. We {reveal} how the description of the whole network can be reduced to the description of just a single superposition mode {(``single-mode model'')}. {In Sec. III,} we show that the regime of interest for us is the initial stage of the dynamics, when strong non-classicality can be obtained. We develop an analytic model for evolution of the Mandel parameter and also show that at this stage the dynamics can be well captured by the linearization approach {for the quantum correction to} the field amplitude. The single-mode model allows to present a simple and illustrative analysis of the PhoG device feasibility providing necessary parameters of the set-up, and {ways} to optimize it.

{In Secs.  IV, V} we analyze dynamics of the system {beyond the simplest single collective mode model. We pay special attention to the two-mode model (Sec. IV) which exhibits a number of nonclassical} features not captured by the single-mode model, such as entanglement generation. After that {in Sec. VI} we proceed to the complete network showing that for the initial stage of dynamics the predictions made {within} the single-mode model are indeed true.  Strong photon-number squeezing is shown to {emerge} for moderate lengths of the CDP circuit with realistic losses and nonlinearity. Also, to assess the impact of specific effects of short-pulse propagation in our system of coupled waveguides, we dynamically
evolved the spectral and temporal properties of the pulse via
the Nonlinear Schrodinger Equation, and analysed their influence on the effectiveness
of the NCL mechanism taking into account the
combined effect of chromatic dispersion, self-phase modulation, and self-steepening of the pulse.

\section{Hierarchy of models}

\begin{figure}[ht]
\includegraphics[width=\linewidth]{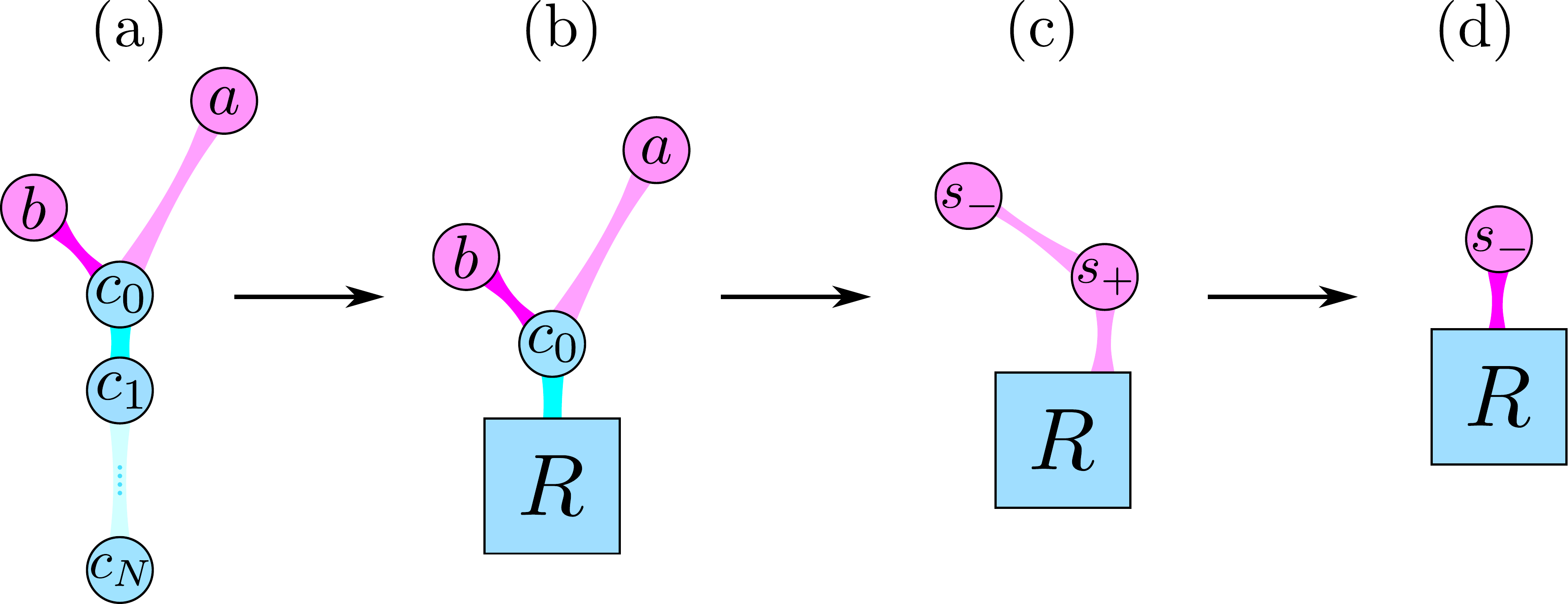}
\caption{{The hierarchy of the PhoG models (a-d). The model (a) corresponds to the set-up shown in the Fig.~\ref{fig:1}(b, c) of coupled single-mode waveguides. The modes $a$ and $b$ interact with the mode $c_0$ coupled also to the ``tail'' of modes $c_{1\ldots N}$. The three-mode model (b) corresponds to the adiabatically eliminated ``tail'' of waveguides retaining only the first waveguide of the tail, which is highly lossy due to direct coupling to the reservoir $R$.  This effectively corresponds to the system presented in \cite{mog2010}. The model (c) shows the ``two-mode model'', the model with two superposition, collective modes $s_{\pm}$ (\ref{modes1}) obtained by eliminating the third mode $c_0$ of the panel (b). Finally, the single-mode model (d) is obtained by adiabatic elimination of one of the collective modes in (c). } }
%MT: I'm happy to make this figure larger if needed.
\label{fig1}
\end{figure}

In the case of a large number of interacting quantum modes, a large number of photons and in the presence of nonlinearity, an exact description of the quantum state dynamics of our system is a formidable task. {To tackle the problem,} we have developed the hierarchical approach to model the photon gun and optimize its structure. The complete CDP scheme {is} reduced to much simpler systems of a few modes {as shown in Fig.~\ref{fig1}}. These systems are much more tractable and can be analyzed, optimization can be carried out, approximations can be verified and then the obtained results {are} {validated by} the approximate numerical analysis of the complete CDP circuit.

The original scheme underlying all the subsequent discussion is depicted in  Figs.~\ref{fig:1}(b,c): two  waveguides $a$ and $b$ are coupled to the third waveguide $c_0$ (but not to each other), and this waveguide $c_0$ is coupled to the ``tail'' of next-neighbour coupled single-mode waveguides $c_j$, $j=1\ldots N$. In our consideration, we assume identical single-mode waveguides described by the corresponding bosonic creation and annihilation operators. We assume that some initial state is created in the network, and then this states evolves. Thus, the dynamics of this system is described by the following master equation for the density matrix $\rho$:
\begin{equation}
\frac{d}{dt}\rho=-i[H,\rho]+\gamma_1\Bigl(\mathcal{L}(a)+\mathcal{L}(b)+\sum\limits_{j=0}^N\mathcal{L}(c_j)\Bigr)\rho,
\label{rN}
\end{equation}
where the Lindbladians $\mathcal{L}$ describing common single-photon loss with the rate $\gamma_1$ are given by $\mathcal{L}(x)y=xyx^{\dagger}-\frac{1}{2}x^{\dagger}xy-\frac{1}{2}yx^{\dagger}x$ for operators $x$ and $y$; the density matrix for all modes is $\rho$ and the Hamiltonian $H=H^{\rm int}+H^{\rm kerr}$ consists of two parts. The first part describes linear interactions between the modes
\begin{equation}
H^{\rm int}=g_aa^{\dagger}c_0+g_bb^{\dagger}c_0+\sum\limits_{j=1}^Ng_jc^{\dagger}_{j-1}c_j +\mathrm{h.c.},
\label{hN}
\end{equation}
{where $g_{a,b}$ are couplings of the modes $a,b$, respectively, to the third mode $c_0$, and $g_j$ is the coupling between mode $c_{j-1}$ and $c_j$. The part $H^{\rm kerr}$ describes} the self-Kerr interaction of each mode, 
\begin{equation}
H^{\rm kerr}=\frac{U}{2}\sum\limits_{\forall x}(x^{\dagger})^2x^2, \quad x=a,b,c_j, \quad j=0,1\ldots N;
\label{kerr}
\end{equation} 
$U$ is the Kerr nonlinear interaction constant.

For large number of waveguides $N$ and large initial number of photons, the complete model Eqs.~(\ref{rN},\ref{hN}) is practically intractable without approximations. However, we are interested in the regime when the ``tail" formed by the waveguides $c_j$, $j=1\ldots N$, can be considered a dissipative reservoir rapidly guiding away field from the mode $c_0$ \cite{natcom,vicencio}.  Assuming that the tail functions as a Markovian reservoir with the decay rate $\gamma_c$, {we arrive at} the three-mode model depicted in Fig.~\ref{fig1}(b) and described by the following master equation for the three-mode density matrix:
\begin{equation}
\frac{d}{dt}\rho_{3}=-i[H_3,\rho_{3}]+\Bigl(\gamma_1\mathcal{L}(a)+\gamma_1\mathcal{L}(b)+\gamma\mathcal{L}(c_0)\Bigr)\rho_{3},
\label{r3}
\end{equation}
where $\gamma=\gamma_1+\gamma_c$, the Hamiltonian is $H_3=H^{\rm int}_3+H^{\rm kerr}_3$ with {$H^{\rm kerr}_3$ as given in Eq.~(\ref{kerr})} and the interaction part is $H^{\rm int}_3=g_aa^{\dagger}c_0+g_bb^{\dagger}c_0+h.c$.

The three mode model of Fig.~\ref{fig1}(b) is still too complicated for exact analysis in case of large number of photons. Further simplification can be reached assuming that the decay rate into the``tail" $\gamma_c$ is large enough to allow for the adiabatic elimination of the mode $c_0$. {Let us introduce} {symmetric and anti-symmetric collective modes}:
\begin{equation}
s_+=\frac{1}{G}(g_aa+g_bb), \quad s_-=\frac{1}{G}(g_ab-g_ba)
\label{modes1}
\end{equation}
with $G=\sqrt{g_a^2+g_b^2}$ and coupling constants $g_{a,b}$ assumed real. {Then adiabatic elimination of mode $c_0$ leads to} the following two-mode master equation \cite{mog2010,mog2011}:
\begin{equation}
\frac{d}{dt}\rho_{2}=-i[H_2,\rho_{2}]+\left(\gamma_1\mathcal{L}(s_-)+(\Gamma+\gamma_1)\mathcal{L}(s_+)\right)\rho_{2},
\label{r2}
\end{equation}
 with $\Gamma=4G^2/\gamma$. {The Hamiltonian $H_2=H^{\rm int}_2+H^{\rm self}_2$ is given by}
\begin{eqnarray}
\nonumber
H^{\rm self}_2=\varsigma_1(n_+^2+n_-^2)+\varsigma_2n_+n_-+\varsigma_3(n_++n_-), \\
H^{\rm int}_2=\varsigma_4(s_+^{\dagger}s_-)^2+\varsigma_5s_+^{\dagger}s_-(n_--n_+-1) +\mathrm{h.c.}
\label{h2}
\end{eqnarray}
where  $n_{\pm}=s_{\pm}^{\dagger}s_{\pm}$ are the photon number operators, and the coefficients read
\begin{eqnarray} \varsigma_1=\frac{U}{2G^4}(g_a^4+g_b^4), \quad
\varsigma_2=\frac{4U}{G^4}(g_ag_b)^2, \quad \varsigma_3=\varsigma_2/4-U/2, \nonumber \\ \varsigma_4=\varsigma_2/4, \quad \varsigma_5=\frac{U}{G^4}g_ag_b(g_a^2-g_b^2). \nonumber \end{eqnarray}  The two-mode model is depicted in Fig.~\ref{fig1}(c).

Finally, {if the state} of the superposition mode $s_+$ decays to the vacuum much quicker than the typical time-scale of the {dynamics of the} superposition mode $s_-$, one arrives to the single-mode model of Fig.~\ref{fig1}(d) described by the following master equation \cite{mog2010,mog2011}:
\begin{eqnarray}
\nonumber
\frac{d}{dt}\rho_{1}=\left(\gamma_1\mathcal{L}(s_-)+\gamma_2\mathcal{L}(s_-^2) + {\gamma_3} \mathcal{L}(n_-s_-)\right)\rho_{1} +\\
 -i[\varsigma_1n_-^2+\varsigma_3n_-,\rho_{1}],
\label{r1}
\end{eqnarray}
where the decay rates {are given by}
\begin{equation}
\gamma_2=\frac{4U^2(g_ag_b)^4}{G^8(\Gamma+\gamma_1)}, \quad \gamma_3=\frac{4U^2(g_ag_b)^2}{G^8(\Gamma+\gamma_1)}(g_a^2-g_b^2)^2.
\label{rates}
\end{equation}
The master equation (\ref{r1}) takes account of the three decay channels for the anti-symmetric collective mode: conventional single photon loss ($\gamma_1$), two-photon loss ($\gamma_2$) and nonlinear coherent loss NCL ($\gamma_3$).  We refer to the decay described by the Lindblad operator $n_-s_-$ as ``nonlinear coherent loss" as the eigenstates of the operator $n_-s_-$ were named ``nonlinear coherent states" \cite{vogel}. The Lindblad operator $n_-s_-$ responsible for NCL can be
considered as the annihilation operator of so-called $f$-deformed harmonic oscillator \cite{manko}.

Note that the unitary part of the master equation, the last term in (\ref{r1}) connected to the self-Kerr interaction, does not manifest itself in the dynamics of the diagonal elements of the density matrix $\rho_1$.

The model (\ref{r1}) is much easier to tackle for large number of photons than the two-mode model (\ref{r2}). Moreover, it allows us to obtain some general conclusions about non-classicality which are valid even for numbers of photons so large that an exact solution becomes unfeasible even for the single-mode model.

\begin{figure}[htb]

\includegraphics[width=\linewidth]{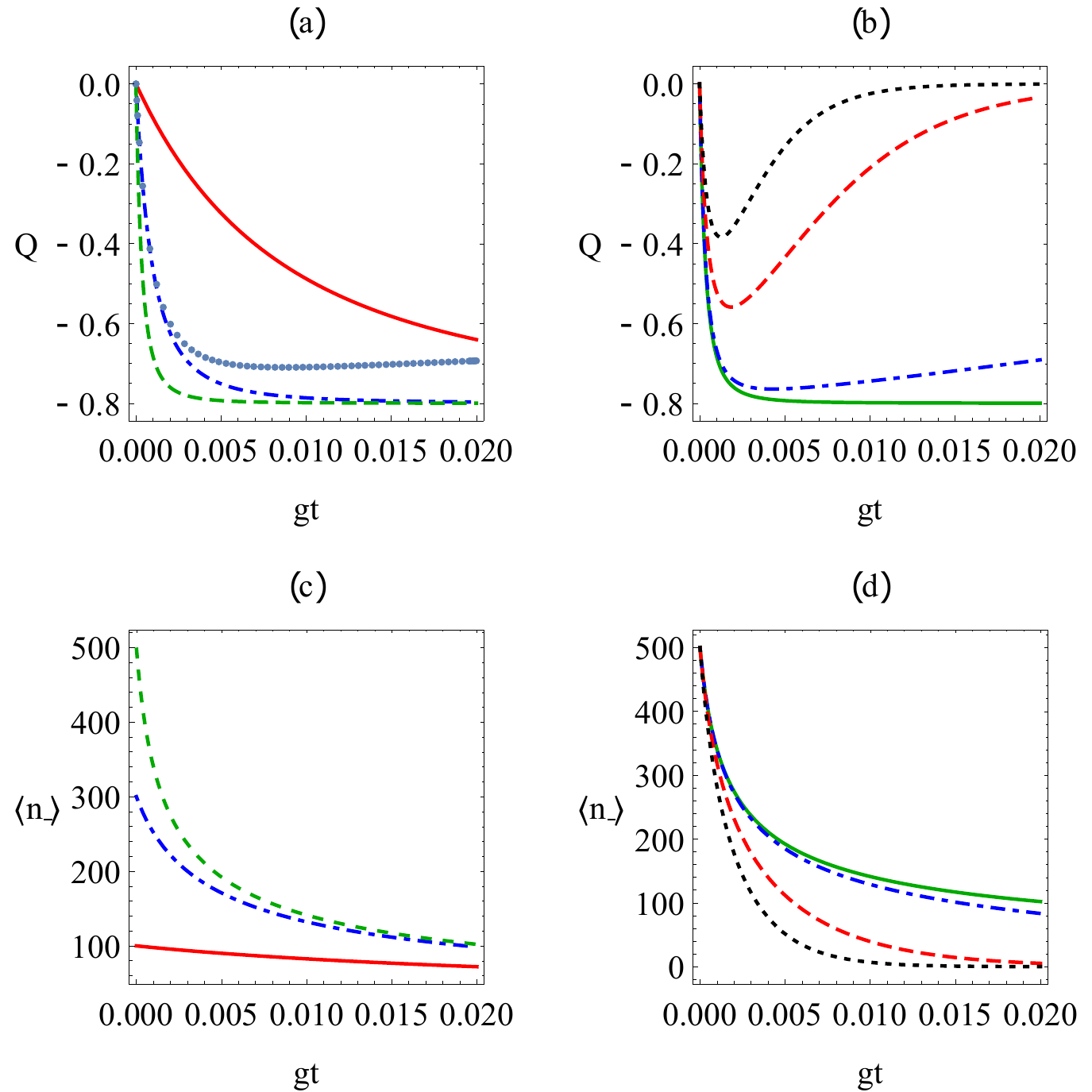}

\caption{The Mandel parameter $Q$  (the panel (a)) and the average number of photons (the panel (c)) for different values of initial number of photons of the input coherent state and no linear loss, $\gamma_1 = 0$. The solid, dot-dashed and dashed lines correspond to $\langle n_-(0)\rangle=100,300,500$ respectively. The line marked with round dots in the panel (a) shows the Mandel parameter for $\langle n_-(0)\rangle = 300$ and $\gamma_2$ increased 100 times (enhanced two-photon loss) compared with that given by Eq.~\eqref{rates}.
The Mandel parameter $Q$  (the panel (b)) and the average number of photons (the panel (d)) for different values of linear loss rate $\gamma_1$ and the initial coherent state with the average number of photons $\langle n_-(0)\rangle=500$. The solid, dot-dashed, dashed and dotted lines correspond to $\gamma_1=0,20g,200g,400g$ respectively. The scaling parameter $g$ here and in the subsequent simulations is taken to be a half of the  third-order nonlinear rate, $U=2g$.
For all the panels the optimal ratio (\ref{optimum}) was taken. Also, it was assumed that the symmetric mode loss rate $\Gamma=432g$. } \label{fig:single_mode_exact}
\end{figure}

\section{Single-mode model}\label{sec:singlemode}

In this section we analyze the simplest single-mode scheme of Fig.~\ref{fig1}(d) to describe the generation of the bright sub-Poissonian state from the semi-classical coherent input, and outline the way to optimize the scheme by choosing the amplitude of the initial coherent state and parameters of the PhoG for realistic CDP circuits with {significant} linear loss and moderate {Kerr} nonlinearity. We establish an existence of universal parameters {$X, Y$, which are determined by effective nonlinearity and thus allow us to characterize the scheme performance even for realistic small optical nonlinearities and, consequently,  very large numbers} of photons, when solving the master equation becomes hardly possible. Finally, we demonstrate practical feasibility of the scheme.

\subsection{Nonlinear loss dynamics}

There are two kinds of nonlinear loss present in the single-mode master equation (\ref{r1}). The first one is the two-photon loss described by the term $\mathcal{L}(s_-^2)$ and the second kind is the NCL described by the term $\mathcal{L}(n_-s_-)$. Both these kinds of losses are able to produce  photon-number squeezing. However, {their} time-scales and asymptotic states are different. The two-photon loss asymptotically leads to a mixture of single-photon and the vacuum states \cite{ezaki,alex}. NCL {asymptotically} leads to the single-photon state \cite{mog2010}.
We show here that for the initial coherent state with large amplitude, $|\alpha|\gg1$ and the assumed equal decay rates, $\gamma_2=\gamma_3$, NCL leads to much faster decay of the average photon number {in the anti-symmetric collective mode $s_-$} than the two-photon loss {and thus to more rapid narrowing of the photon number distribution in $s_-$}.

The degree of photon-number distribution squeezing can be conveniently described by the Mandel parameter \cite{Davidovich1996}{, which for the superposition mode $s_-$ reads:}
\begin{equation}
Q=\frac{\langle(s_-^{\dagger})^2s_-^2\rangle}{\langle n_-\rangle}-\langle n_- \rangle.
\label{q}
\end{equation}
Mandel parameter $Q = {-}1$ corresponds to the perfect squeezing (Fock states), zero is for coherent states.
This parameter can be inferred from dynamics of only the diagonal elements of the density matrix (\ref{r1}) given by the following equation
\begin{eqnarray}
\nonumber
\frac{d}{dt}\rho_n=-(\gamma_1n+\gamma_2n(n-1)+\gamma_3n(n-1)^2)\rho_n+\\
\label{diag}
(\gamma_1(n+1)+\gamma_3(n+1)n^2)\rho_{n+1}+\\
\nonumber
\gamma_2(n+1)(n+2)\rho_{n+2},
\end{eqnarray}
where $\langle n|\rho_{1}(t)|n\rangle=\rho_n$ are the density matrix elements in the Fock-state basis.

Note that for the symmetric case, $g_a=g_b$, the NCL decay rate $\gamma_3$ is zero. For asymmetric coupling, $g_a\neq g_b$,   we fix the parameter $g_a$ and start varying the parameter $x=g_b/g_a$ for $x\in[0,1]$. The maximal value of the rate $\gamma_3$ is then given by the condition $d\gamma_3(x)/dx=0$ and
leads to the following  result for the ratio of the interaction constants {to maximize NCL}:
\begin{equation}
\left ( g_b/g_a \right )_{\rm opt} = \sqrt{2}-1.
\label{optimum}
\end{equation}
{For this optimal coupling ratio}, we have
\begin{equation}
\gamma_3=\frac{U^2}{4(\gamma_1+\Gamma)}
\label{gopt}
\end{equation}
In further considerations, we adopt this ratio for estimations.

In Fig.~\ref{fig:single_mode_exact} examples of the Mandel parameters and average photon number dynamics are given for different numbers of photons (several hundred) of the initial coherent state.  Fig.~\ref{fig:single_mode_exact} {reveals} a number of important features {in the dynamics of the antisymmetric collective mode}:
\begin{enumerate}

\item The Mandel parameter {rapidly} decreases {at} the initial stage of the dynamics. Only a small percentage of photons is lost when quite considerable {negative} values of the Mandel parameter are reached (less than $15\%$ to have {significant photon-number squeezing with} $Q<-0.25$, see Figs.~\ref{fig:single_mode_exact}(c,d)).

\item The practically reachable values of the Mandel parameter are approximately limited  to $Q=-0.8$ (see also Ref.~\cite{mog2013}). This minimal value is shifted to smaller interaction length by {increasing the average number of photons in} the initial state.

\item Dynamics due to nonlinear coherent loss (NCL) occurs much faster than dynamics due to the two-photon loss. {As seen} in Fig.~\ref{fig:single_mode_exact}(a), {even hundredfold increase in} the two-photon decay rate, $\gamma_2$, does not influence much the initial stage of the $Q$-parameter dynamics. In general, the ratio between these two time scales is determined by the ratio between the coupling constants, $g_b/g_a$: $\gamma_3 = \gamma_2 \left ( 1 - \left ( g_b/g_a \right )^2 \right ) / \left ( g_b/g_a \right )^2$ (see Eq.~(\ref{rates}) and text around Eq.~(\ref{optimum})).

\item NCL can outrun linear loss. Fig.~\ref{fig:single_mode_exact}(b) {shows} that even large linear loss weakly affects the initial stage of the dynamics.

\item {Consider now the case of no linear loss}. For different large initial numbers of photons, after some time the average numbers of photons tends to {largely the same stationary value (Fig.~\ref{fig:single_mode_exact}(c))}. It is a signature behaviour of NCL \cite{mog2010}.

\end{enumerate}

\begin{figure}
\includegraphics[width=0.75\linewidth]{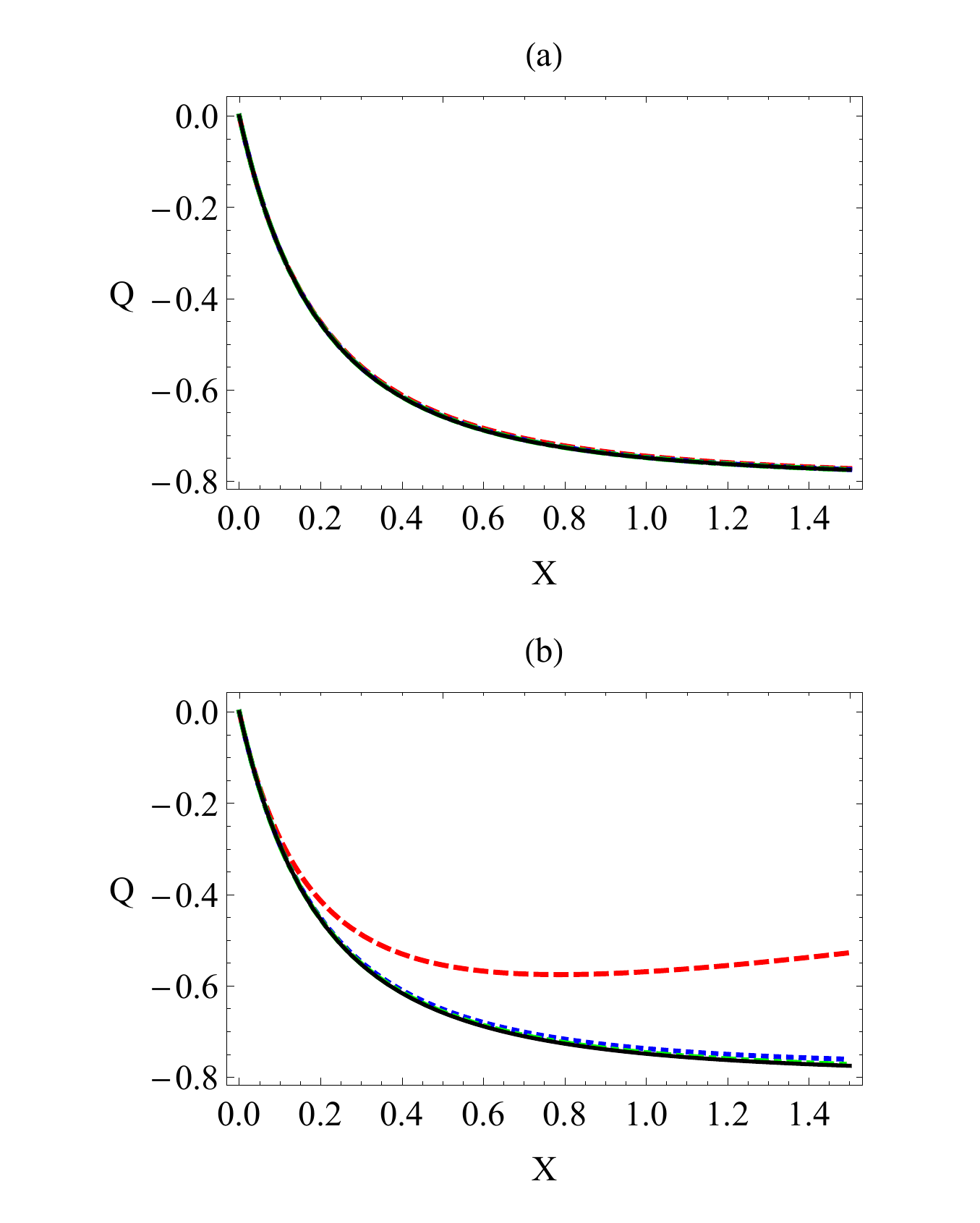}
\caption{The dependence of the Mandel parameter $Q$ on the parameter $X$ of Eq.~(\ref{x}) for different initial average numbers of photons $\langle n_-(0)\rangle$ in absence of linear loss, $\gamma_1=0$,~(a) and for significant linear loss, $\gamma_1=10g$,~(b). Dashed, dotted and dotdashed lines were obtained by solving Eq.~\ref{diag} and correspond to $\langle n_-(0)\rangle=100, 400, 900$ respectively. Solid line corresponds to the analytical approximation given by Eq.~(\ref{mandel}). Other parameters are the same as for Fig.~\ref{fig:single_mode_exact}. The lines merge together and  are almost indistinguishable for $\gamma_1 = 0$. The line for $\langle n_-(0)\rangle = 900$ merges with approximate solution~\eqref{mandel} even for $\gamma_1 = 10g$.}\label{fig3}
\end{figure}

\subsection{Universal parameters}

{As we} have shown, the two-photon loss only weakly influences the dynamics of photon number squeezing. Also, large photon-number squeezing is generated at the initial stage of dynamics when only relatively small number of photons is lost. These two facts allow us to introduce two dimensionless parameters allowing an estimation of the feasibility of the scheme. These two parameters are
\begin{equation}
X=\gamma_3 (\langle n_-(0)\rangle)^2 t_{fix}, \quad Y=\frac{\gamma_1}{\gamma_3\langle n_-(0)\rangle^2},
\label{x}
\end{equation}
where $t_{fix}$ is the fixed interaction time (defined by the length of the PhoG). The parameter $X$ defines the reachable value of the Mandel parameter, $Q$, in absence of the linear and two-photon loss, and the parameter $Y$ defines the tolerable level of the linear loss when the value of $Q$ is still defined by $X$. 
	
In terms of parameter $Y$, one has to have $Y<1$ for optimal $Q$ for a given $X$, Fig.~\ref{fig3}~(a). Of course, even larger values of $Y$ still allow for considerable $Q$. However for that, larger values of $X$ are required.

Such a dependence of photon number squeezing on the universal parameters can be explained by simple analytic considerations. For large number of photons of the initial coherent state, the initial photon number distribution can be with high precision described by the Gaussian. Let us assume that for initial stage of the dynamics the distribution can still be described by the Gaussian
\begin{equation}
\rho_n(t)=\frac{1}{\sqrt{2\pi}\sigma(t)}\exp\left\{-\frac{(n-\langle n_-(t)\rangle)^2}{2\sigma(t)^2}\right\}
\label{gauss}
\end{equation}
where $\sigma(t)=(Q(t)+1)\langle n_-(t)\rangle$. In the Appendix  A it is shown that Gaussianity of the photon-number distribution  holds very well for $\langle n_-(t)\rangle\gg1$.  Assumption of a smooth photon number distribution slowly changing with $n$ (like the one given by Eq.(\ref{gauss}) for large $\langle n_-(0)\rangle$) allows to introduce a continuous variable $n$, to make an approximation
$\rho_{n+m}\approx \rho_{n}+m\frac{d}{dn}\rho_{n}+m^2\frac{1}{2}\frac{d^2}{dn^2}\rho_{n}$, and to calculate averages as
\[\langle n_-(t)^m\rangle=\int\limits_0^{+\infty}dn\rho_nn^m.
\]
Thus, one can obtain the following equation for the Mandel parameter 
\begin{equation}
\frac{d}{dt}Q\approx -\gamma_1Q-2\gamma_2\langle n_-\rangle(3Q+1)-\gamma_3\langle n_-\rangle^2(5Q+4).
\label{qeq}
\end{equation}
Neglecting linear and two-photon losses, in the continuous approximation for Eq.~(\ref{diag}) one obtains 
that $\langle n_-(X)\rangle\approx\langle n_-(0)\rangle/\sqrt{1+2X}$ (see the Appendix A). For the Mandel parameter we get:
\begin{equation}
Q(X)\approx -\frac{4}{5}\left(\frac{1}{(1+2X)^{{5}/{2}}}-1\right).
\label{mandel}
\end{equation}

This situation is illustrated in Fig.\ref{fig3}. The panel \ref{fig3}(a) shows that in the absence of linear loss neither the initial number of photons nor the  interaction time by themselves are the parameters defining possible $Q$. Practically, it depends only on $X$.  The black solid line in the panel \ref{fig3}(a) shows the result provided by the approximation (\ref{mandel}) for several initial states with hundreds of photons. It is obvious that Eq.(\ref{mandel}) very closely reproduce the exact solution of Eq.~(\ref{diag}). Eq.~(\ref{diag}) also shows why the influence of two-photon losses is so small for $\gamma_2$ close to $\gamma_3$: the time scales are determined by $\propto \gamma_2 n^2$ for two-photon loss and $\propto \gamma_3 n^3$ for NCL. For large photon numbers and $\gamma_2 \approx \gamma_3$, the NCL obviously dominates. 

The panel \ref{fig3}(b) shows an influence of linear loss (described by the parameter $Y$) on the reachable value of the Mandel parameter.  For a given linear loss of $\gamma_1=10g$, for $\langle n(0)\rangle=900$ one has $Y<1$, and the curve describing the achievable values of the Mandel parameter (dash-dotted line) is practically coinciding with the approximation given by Eq. (\ref{mandel}). For small $Y$ any influence of the linear loss is indeed small. For smaller initial number of photons, 
$\langle n(0)\rangle=100$, when $Y>1$, one has considerable difference with analytic predictions  (dashed line in Fig.\ref{fig3}(b)).

\subsection{Feasibility}

Universal parameters $X$ and $Y$ are decisive to ensure the optimal performance of PhoG as a deterministic source of sub-Poissonian light. For PhoG design we have a ``rule-of-thumb'': we should aim to make $X$ as large as possible, and $Y$ as small as possible. In this section we demonstrate that it is  feasible to reach $X>0.3$ and $Y<1$  for realistic waveguide structures.

First of all, we connect the nonlinear interaction constant $U$ in the expressions (\ref{rates}) with quantities commonly used for description of the Kerr-nonlinearity in waveguide structures. We assume that the pulse of the finite length propagates through the waveguide, and the pulse length $L_{\rm eff}$ is much smaller than the waveguide length; we introduce the propagation length as $z=ct_{\rm fix}/n_{\rm eff}$, where $n_{\rm eff}$ is the effective refractive index of the waveguide mode. Thus, the nonlinear interaction constant in our model can be written in the following way (see, for example, Refs.~\cite{drummond}):
\begin{equation}
U=2\hbar\omega\frac{\omega}{V_{\rm eff}}\frac{n_2}{n_{\rm eff}},
\end{equation}
where $n_2$ is the nonlinear refractive index and $V_{\rm eff}$ is the mode volume.
For our model, we assume $V_{\rm eff}\approx A_{\rm eff}L_{\rm eff}$, where $A_{\rm eff}$ is the effective area. The effective area
is defined as
\begin{equation}
A_{\rm eff}^{-1}=\int\int d^2{\vec s} \ |E(\vec s)|^4,
\label{a}
\end{equation}
where $E(\vec s)$ is the normalized transversal field profile,
\[
\int\int d^2{\vec s} \ |E(\vec s)|^2=1.
\]
Here we do not consider influence of losses on the nonlinear interaction constant \cite{coso}.

In a fiber-like waveguides it is customary to introduce the nonlinear fiber coefficient:
\begin{equation}
\bar\gamma^{\rm NL}=\frac{\omega}{c}\frac{n_2}{A_{\rm eff}}\Rightarrow U=2\frac{\hbar\omega}{T_{\rm eff}n_{\rm eff}}\bar\gamma^{\rm NL},
\end{equation}
where $T_{\rm eff}$ is the pulse time-duration in vacuum.

Now let us consider several waveguide arrangement with realistic parameters.

\subsubsection{Large waveguides in bulk glass}\label{sec:large_waveguides}

Let us take coupling parameters as used in the CDP circuits in Ref.~\cite{natcom}, $g_a\approx 200 m^{-1}$.  The decay rate of the $c_0$ mode Fig.~\ref{fig1} can be approximated by $\gamma_c\approx 4\sqrt{g_a^2+g_b^2}$, which for optimal coupling ratio results in $\Gamma\approx 216.5$m$^{-1}$ (decay rate for the symmetric collective mode $s_+$, Fig.~\ref{fig1}). For intended linear loss of about $0.5$~dB/cm, we have $\gamma_1\approx 11.5 $m$^{-1}$, and $\gamma_3\approx 0.0011$m$^{-1} \times U^2$. The dimensions of the waveguides  as used in CDP circuits in Ref.~\cite{natcom} are approximately $4 \mu$m$\times 4 \mu$m$ $.  For a conservative estimate, let us assume an effective modal area an order of magnitude larger than typical for single-mode fibers with the similar core dimensions around $1000$ nm  (see, for example, Corning® HI 1060 fiber with the mode-field diameter of about 6.2 $\mu$m at $1060$ ~nm), $A_{\rm eff}=300\mu m^2$ (which would be more than three times larger even for typical modal areas at 1550 nm).
Thus, taking $n_2=3\times 10^{-18}$W$^{-1}$m$^2$ and $n_{\rm eff}=2.59$ typical for IG2 glass \cite{ig2}, for a $100$~fs pulse at $1060$~nm we have $\gamma_3\approx 8\times 10^{-18}$m$^{-1}$. Condition $Y=1$ of Eq.~(\ref{x}) requires energy levels of $1.2 \times 10^{9}$ photons per pulse or $224$pJ per pulse. The feasibility of these energy levels in the context of our set-up is confirmed by the recent work \cite{robert2018}, where femtosecond pulses with energies more than  $10$~nJ  at $1030$~nm has been used for writing waveguides. Using all these parameters, for a $3$~cm waveguide we have $X\approx 0.33$, which is sufficient to reach high degree of photon-number squeezing $Q<-0.5$. As follows from the condition (\ref{x}), for lower effective modal area one would need proportionally lower energies per pulse. For $A_{\rm eff}=30\mu $m$^2$ only $25$pJ per pulse is sufficient.

\subsubsection{Fiber-like systems}

 Now let us use for the PhoG set-up the parameters from the recent fiber-based interferometric scheme Ref.~\cite{aruto}  for sub-Poissonian light generation. For the used polarization maintaining fiber (Nufern, HP-780) one has ${\bar\gamma}^{\rm NL}=8.51\times 10^{-3}$W$^{-1}$m$^{-1}$ for $808$~nm. Using $n_{\rm eff}=1.45$, we arrive at $\gamma_3\approx 9.2 \times 10^{-19}$m$^{-1}$. It is nearly an order of magnitude lower than for the bulk waveguides. However, much lower loss
of $3.5$~dB/km implies that the condition (\ref{x}) is easier to satisfy. For that, $3 \times 10^7$ photons per pulse (or less than $10$pJ per pulse) are needed. If $2 \times 10^{9}$ photons per pulse are used, $10$cm of the fiber are required to reach $X=0.37$ and squeezing of about $Q<-0.5$ (which is more or less in agreement with results obtained in Ref.~\cite{aruto}).

\subsubsection{Silicon nanowires}

Finally, let us consider silicon nanowires arrangement as a possible platform for PhoG generators. Si-nanowire waveguides have typically very high nonlinearity and loss, and small lengths. We assume loss of $5$~dB/cm and moderate nonlinearity, ${\bar\gamma}^{\rm NL}=300$W$^{-1}$m$^{-1}$ with other parameters being the same as above \cite{sato}. Notice that even for such loss level,  $\Gamma$ is larger than $\gamma_1$, where $\gamma_1\approx 115$~m$^{-1}$. For $n_{\rm eff}=3.5$ and wavelength of $1064$~nm, we obtain $\gamma_3\approx 7.75 \times 10^{-11}$m$^{-1}$, which is much larger than for all the previously considered cases. To satisfy $Y=1$, we require only $1.3 \times 10^6$ photons per pulse or less than $0.3$pJ pulse energy. For this number of photons a $3$~mm waveguide would be sufficient to achieve $X=0.345$ and $Q<-0.5$.

\subsection{Linearization}\label{sec:single_mode_linear}

Here we demonstrate that, over the timescales when large photon number squeezing occurs, at the initial stage of dynamics the evolution of the Mandel parameter can be adequately described using a linearization approach.  This approximation is typical for consideration of nonlinear waveguide systems. Essentially, it is a linearization on a quantum correction to the creation/annihilation operators, $s_-=S_-+\delta s_-$, where the quantity $S_-=\langle s_-\rangle$ is the classical amplitude, $\delta s_-$ are the quantum fluctuations which obey the same commutator as $s_-$, and $\langle \delta s_- \rangle = 0$. The resulting system of equations allows us to find quantities $\langle s_-\rangle$, $\langle s_-^\dagger\rangle$, $\langle s_-^2\rangle$,  $\langle s_-^{\dagger}s_-\rangle$ and $\langle s_-^{\dagger 2}\rangle$ and circumvents the computational complexity required in the case of exact solution for large photon numbers in a large number of modes.

\begin{figure}[hb]
\centering
\includegraphics[width=0.75\linewidth]{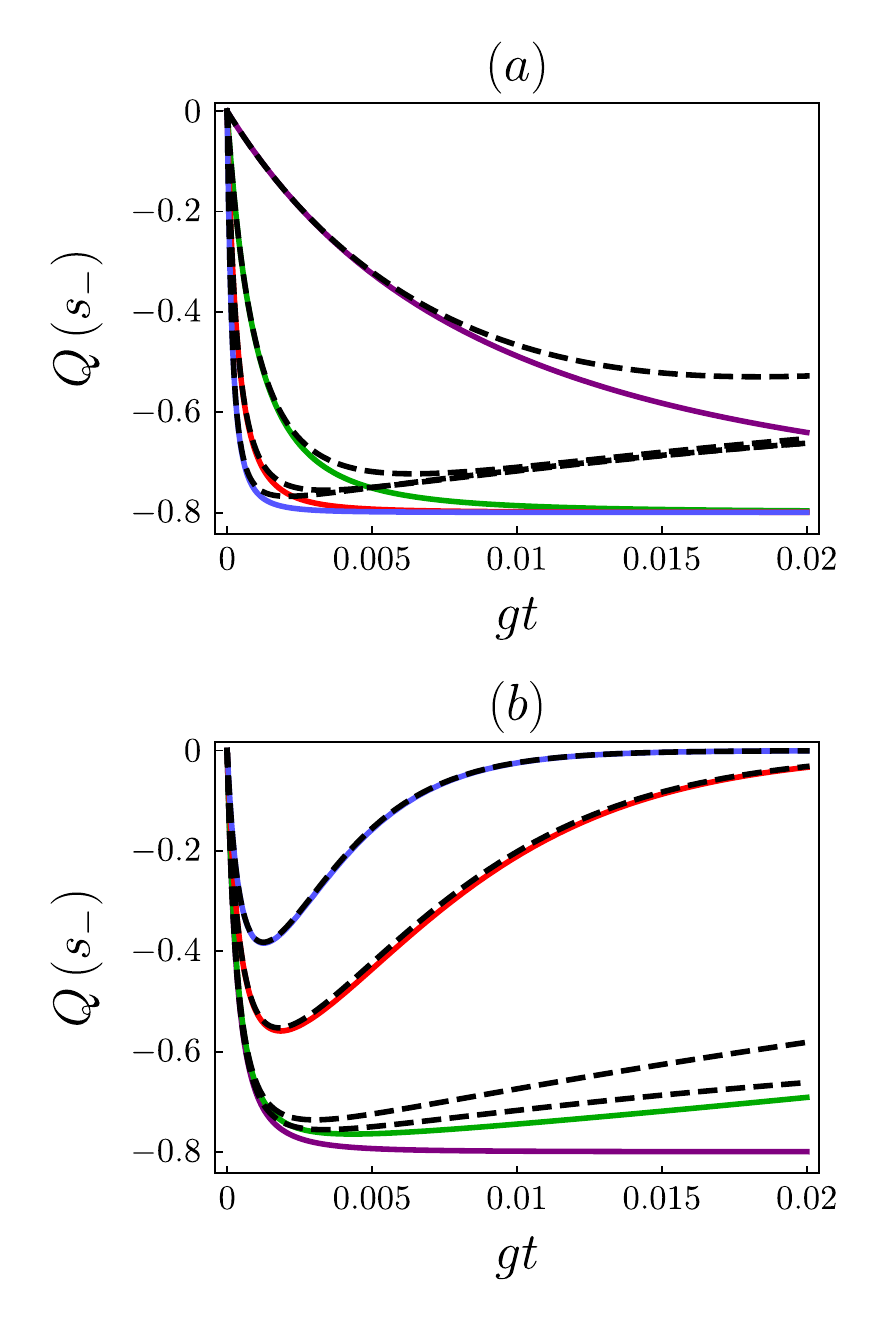}
\caption{\label{fig:Qlinvcompare} 
Evolution of the Mandel $Q$ parameter is accurately predicted by the linearization method over the initial stages of evolution, and $|Q|$ is underestimated at later stages. {Dashed}: linearized. {Solid}: exact, as in Fig.~\ref{fig:single_mode_exact}. $U = 2g$ and $\Gamma = 432g$. (a) Initial photon number $\langle n_-\left(0\right)\rangle = 100, 300, 500, 700$ (top to bottom), with $\gamma_1 = 0$. (b) $\langle n_- \left(0\right)\rangle = 500$, $\gamma_1 = 0, 20g, 200g, 400g$ (bottom to top). With realistic linear loss rates $\gamma_1$ our approximation remains accurate even in the late stages of evolution. Both graphs use the same time scaling $g t$ as Fig.~\ref{fig:single_mode_exact}}
\end{figure}

Our starting point is Eq.~(\ref{r1}), from which we may derive the following system of coupled equations: 
\begin{align}\label{eqn:expectations}
% a
&\partial_t\langle s_-\rangle =  c_1  \langle s_-\rangle + c_2\langle s_-^\dagger s_-^2\rangle \notag - \frac{\gamma_3}{2} \langle s_-^{\dagger 2} s_-^3 \rangle \notag \\
% ad
&\partial_t\langle s_-^\dagger \rangle =  c_1^*\langle s_-^\dagger\rangle + c_2^* \langle s_-^{\dagger 2} s_- \rangle - \frac{\gamma_3}{2} \langle s_-^{\dagger 3} s_-^2\rangle \notag \\
% aa
&\partial_t \langle s_-^2 \rangle = c_3\langle s_-^2 \rangle + c_4 \langle s_-^\dagger s_-^3 \rangle - \gamma_3 \langle s_-^{\dagger 2} s_-^4 \rangle \notag \\
% adad
&\partial_t \langle s_-^{\dagger 2} \rangle = c_3^* \langle s_-^{\dagger 2} \rangle + c_4^*\langle s_-^{\dagger 3} s_-\rangle  -\gamma_3 \langle s_-^{\dagger 4} s_-^2\rangle \notag \\
% ada
&\partial_t\langle s_-^\dagger s_- \rangle = - \gamma_1 \langle s_-^\dagger s_-\rangle + c_5 \langle s_-^{\dagger 2} s_-^2\rangle - \gamma_3 \langle s_-^{\dagger 3} s_-^3\rangle
\end{align}
with coefficients $c_1=\left(-\gamma_1/2 + i \varsigma_1 + i \varsigma_3\right)$, $c_2=\left(\gamma_2 - \gamma_3 + 2 i \varsigma_1\right)$, $c_3=\left(-\gamma_1 - \gamma_2 - \gamma_3 + 4 i \varsigma_1 + 2 i \varsigma_3\right)$, $c_4=\left(-2\gamma_2 - 5 \gamma_3 + 4 i \varsigma_1\right)$ and $c_5=\left(-2\gamma_2 - \gamma_3\right)$. 
\par
This system of equations is not closed, and so cannot yet be solved. To proceed, we must perform the linearization procedure in order to reduce  operator products in Eq.~(\ref{eqn:expectations}) to at most second-order, thereby closing the system of equations and allowing us to obtain its solution. The linearization  can be done in a standard manner using the cumulant expansion (see Appendix~\ref{appendix:linear}). The linearized forms of Eqs.~(\ref{eqn:expectations}) are excplitly shown in Appendix~\ref{appendix:single_mode_linear}.

%Retaining the correlations up to the second order, we approximate averages of the operator products for arbitrary operators $A, B, C, D$ in the following way (Appendix~\ref{appendix:linear}):
%\begin{align}\label{eqn:linear_approximation}
%\langle A B C \rangle \approx &\langle A \rangle \langle B C \rangle + \langle B \rangle \langle A C \rangle + %\langle C \rangle \langle A B \rangle \notag \\
%&- 2 \langle A \rangle \langle B \rangle \langle C \rangle  \notag \\
%\langle A B C D \rangle \approx &\langle A B \rangle \langle C D \rangle + \langle A C \rangle \langle B D \rangle %+ \langle A D \rangle \langle B C \rangle \notag \\
%- 2 \langle A \rangle \langle B \rangle \langle C \rangle \langle D \rangle.
%\end{align}
%In Appendix~\ref{appendix:linear} the linearized forms of Eqs.~(\ref{eqn:expectations}) are explicitly shown.

The solution of the linearized forms of Eq.~(\ref{eqn:expectations}) are used to approximate $Q$, and in Fig.~(\ref{fig:Qlinvcompare}) this result is compared to the exact method from the previous section. The linearization approximation (dashed lines) accurately predicts the evolution of $Q$ over the initial stages of evolution, and actually underestimates $|Q|$ in the later stages, although including realistic $\gamma_1$ allows the approximation to remain accurate as the nonclassical output state is pushed towards the vacuum.

Since when $\gamma_1 = 0$ the linearization approximation remains accurate over the timescales of interest (see the panel Fig.~\ref{fig:Qlinvcompare}(a)), and since the presence of realistic loss makes the approximation increasingly accurate (see the panel Fig.~\ref{fig:Qlinvcompare}(b)), we may confidently apply the linearization approach over the parameter regimes of physical interest -- short times and realistic loss -- even in the case where a fully quantum treatment would be intractable, i.e. large $\langle n\left(0\right)\rangle$ or large number of modes.

\section{The symmetric two-mode model}

In the previous section, we have shown that two-photon loss is far less efficient in producing photon-number squeezing than NCL. However, it is worth pointing out that even symmetric system exhibiting only two-photon loss can be of use. To show it,  let us recourse to the two-mode model of Fig.~\ref{fig1}(c).  First of all, even a symmetric system is able to produce entangled states of waveguide modes asymptotically, when the state of the symmetric superposition mode $s_+$ is the vacuum (Sec.~\ref{section:modal_entanglement}). 
The second and rather non-trivial feature of the symmetric system is the possibility of producing few-photon entangled states by driving the rapidly decaying collective mode $s_+$ (Sec.~\ref{section:photon_pairs_generation}). This scheme can be also used for probabilistic generating  of single photons. 

\subsection{Modal entanglement}\label{section:modal_entanglement}

Generally, the state produced by the two-photon loss is quite far from Gaussian (asymptotically, the initial bright coherent state is driven by the two-photon loss toward the superposition of the vacuum and single-photon state \cite{ez1,ez2}). However, one can still obtain a lower bound on the entanglement between modes from the covariance matrix for modes $a, b$: $\sigma_{kl}=\frac{1}{2}\langle d_kd_l+d_ld_k\rangle-\langle d_k\rangle\langle d_l\rangle$, with the vector $\vec{d}=\left(1/\sqrt{2}\right) \left[a+a^\dagger,i\left(a^\dagger-a\right),b+b^\dagger,i\left(b^\dagger-b\right)\right]$ \cite{Adesso2007,Weedbrook2012}. This bound is provided by the (Gaussian) logarithmic negativity $\mathcal{N}$ \cite{Weedbrook2012, Adesso2007}, which quantifies violation of the PPT~criterion \cite{Simon2000}. The logarithmic negativity $\mathcal{N}$ is defined as 
$\mathcal{N}=\max{\left\{0,-\log\lambda\right\}}$, where $\lambda$ is  the smallest symplectic
eigenvalue of the partially transposed matrix $\sigma_{kl}$. Elements of this matrix are provided by the linearized approach considered in the previous section.

For example, for $\langle a^\dagger a \left(0\right)\rangle = \langle b^\dagger b \left(0\right)\rangle =2500$, $U  = 2g$, $\gamma_c = 15g$, $\gamma_1 = 11.5g$ and symmetric coupling $g_a = g_b = 60g$, the system evolves to logarithmic negativities $\mathcal{N} \approx 1.25$ between modes $a$ and $b$ within $0.01$~$gt$, while modes $a$ and $b$ each contain $18$~photons. %The corresponding maximally-entangled Gaussian states (two-mode squeezed vacuum states) with equivalent photon numbers have logarithmic negativites of $4.1$ respectively.
Thus symmetric PhoG generates entanglement between modes $a$ and $b$ (though is unable to produce large photon-number squeezing in the initial stages of the dynamics).

\subsection{Photon pairs generation}\label{section:photon_pairs_generation}

\begin{figure}[h]
\includegraphics[width=0.8\linewidth]{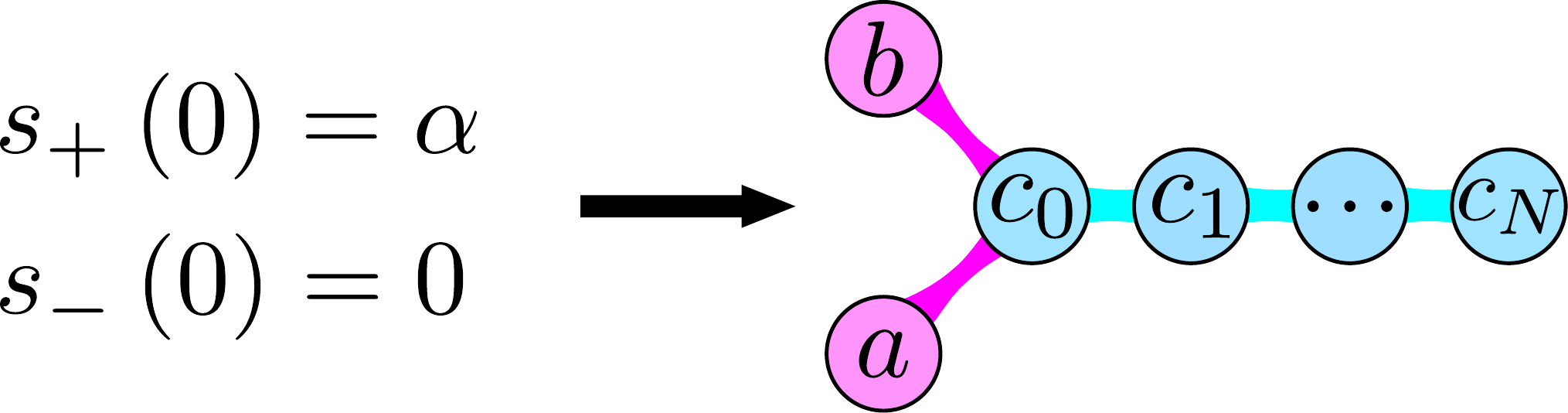}
\caption{ The two-mode PhoG model: symmetric CDP circuit with initial excitation in symmetric collective mode $s_+$.}
\label{fig-symPhoG}
\end{figure}

Despite the quick decay of the symmetric mode $s_+$, it still can create a non-classical state in the mode $s_-$.  Indeed, for the symmetric case of Fig.~\ref{fig-symPhoG}, $g_a=g_b\equiv g$, we have $\varsigma_5=0$ and the Hamiltonian part describing interaction between superposition modes is
\begin{eqnarray}
H^{int}_2=\varsigma_4(s_+^{\dagger}s_-)^2 +\mathrm{h.c.}
\label{hs}
\end{eqnarray}
The Hamiltonian $H^{int}_2$ for the case describes the four-wave mixing process of transferring two photons of initially excited decaying mode $s_+$
to the initially empty mode $s_-$. Thus the symmetric PhoG of Fig.~\ref{fig-symPhoG} can be designed to reproduce the four-wave mixing process which is known to allow for the generation of entangled photon pairs \cite{kumar,tak}. But notice that in our case there is no problem of separating the generated state from the pump. For sufficiently long PhoG the driving excitation of the mode $s_+$ completely decays leaving photons only in the mode $s_-$.

\begin{figure}[h]
\includegraphics[width=\linewidth]{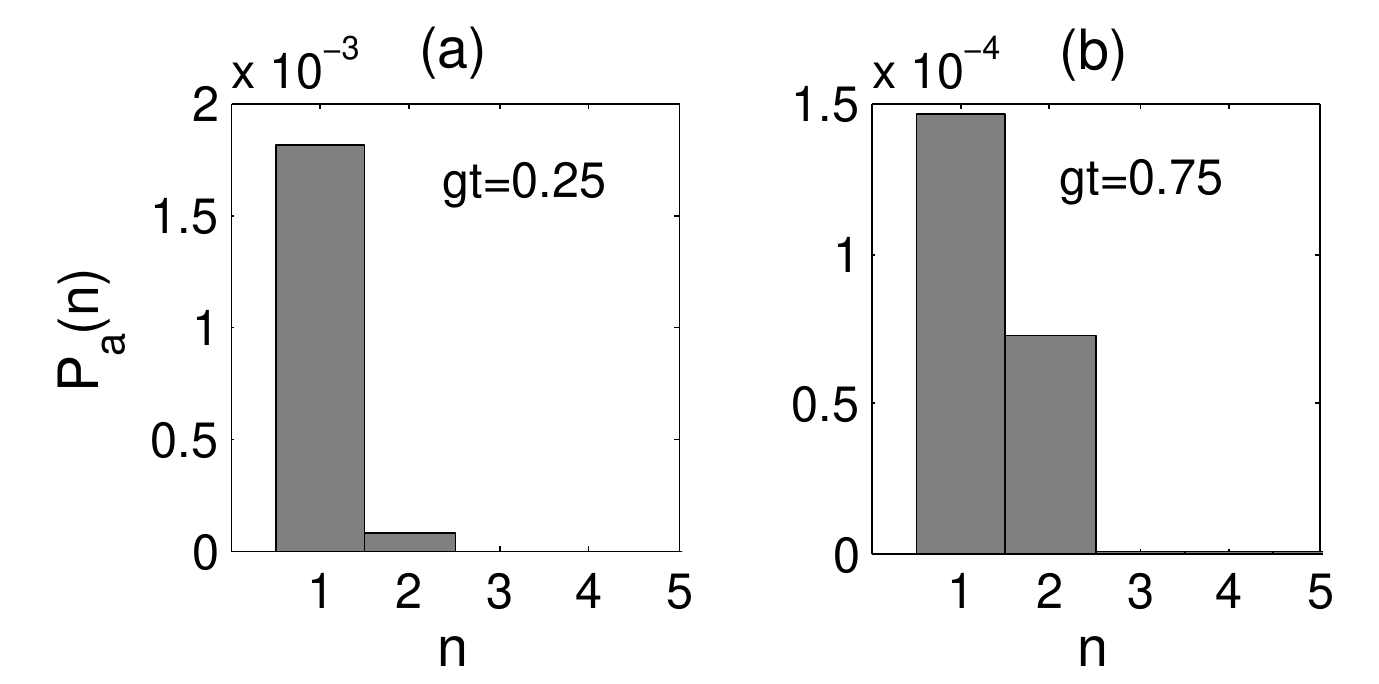}
\caption{$P(n)$, the n-photon probabilities, in the mode $a$ for different propagation times. The symmetric PhoG is taken, $g_a=g_b=g$. Weak coherent states with amplitudes $\alpha_a=0.5$; $\alpha_b=0.5$ are assumed as initial states of the modes $a$ and $b$.  Here zero-photon components are not shown. Notice that here the scaling constant $g$ is taken to be equal to the Kerr nonlinear constant, $U=g$; the loss in the collective symmetric mode is $\gamma_c=5g$, linear loss rate is zero.}
\label{fig6}
\end{figure}

 The process of the pair generation dynamics with weak coherent initial states is illustrated in Fig.~\ref{fig6} with the photon-number distribution of the mode $a$. Initial coherent states with equal amplitudes are excited in the waveguides $a$ and $b$ (corresponding to $s_+, s_-$ as shown in Fig.~\ref{fig-symPhoG}). For small propagation lengths, the remnant of the initial excitation dominates (Fig.~\ref{fig6}(a)). However, it quickly decays and for larger times predominantly two photons in the anti-symmetric mode are observed (Fig.~\ref{fig6}(b)). Just two photons in this mode corresponds to the following entangled superposition of photons in the waveguides $a$ and $b$ at the output of the PhoG:
\begin{eqnarray}
|s_-\rangle^{\rm out}|s_+\rangle^{\rm out}=|2\rangle_-\ |0\rangle_+=  \label{state} \\
\frac{1}{2}\left(|2\rangle_a|0\rangle_b+|0\rangle_a|2\rangle_b-\sqrt{2}|1\rangle_a|1\rangle_b\right). \nonumber
\end{eqnarray}
One can see the photon-number distribution corresponding to single- and two-photon part of the generated state (\ref{state}) in panel  Fig.~\ref{fig6}(b). The ratio of the single-photon component to the two-photon component is approximately 2.05 for $gt=0.75$ .
Notice that by increasing amplitudes of the initial states, a squeezed vacuum state is created in the mode $s_-$.  

Another important difference with the conventional four-wave mixing pair-generating schemes \cite{kumar,tak} is that here we have the two entangled photons in two spatially separated modes. As can be seen in Ref.~\cite{natcom}, at the PhoG output, the spatial distance between the modes exceeds considerably the mode cross-section.

\begin{figure}
    \centering
    \includegraphics[width = \linewidth]{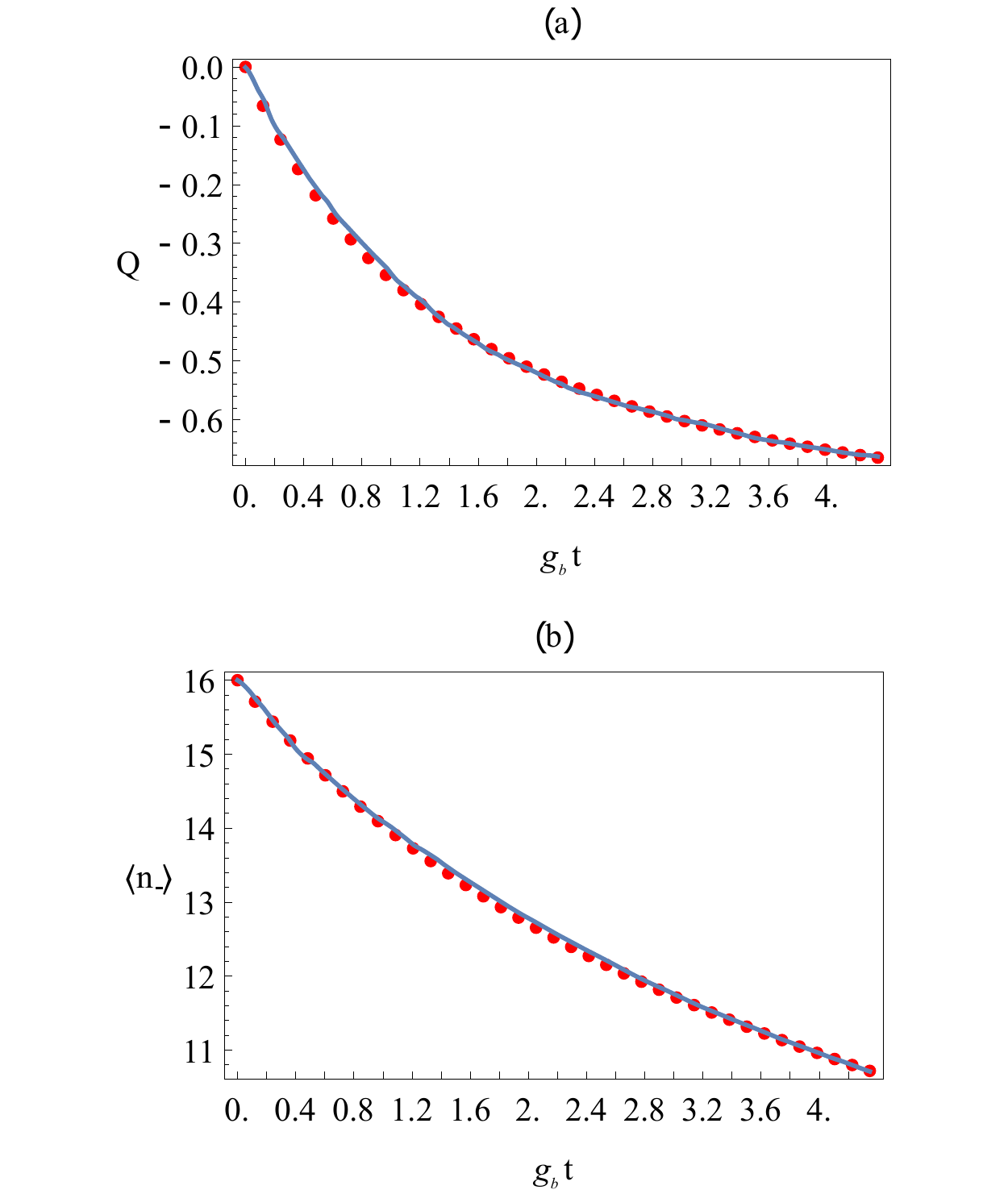}
    \caption{Evolution of Mandel parameter~(a) and mean photon number~(b) of the antisymmetric mode $s_{{-}}$ computed using the three modes model (solid lines) and the single mode model (dots) for the following parameters: $g_b/g_a=\sqrt{2}-1$, $U = 0.012 g_b$, $\gamma_c = 6.04 g_b$, $\gamma_1 = 0$.}\label{fig:ThreeModes_vs_SingleMode}
\end{figure}

\section{Three-mode model}

As has been seen in the previous Sections, analysis of the single-mode model allows us to give simple and straightforward prescriptions for the practical realizations  of the PhoG. The question is how close predictions given by the single-mode model approach the results for more complicated modal arrangements.  

In this Section we analyze dynamics of a three-mode PhoG schematically shown in Fig.~\ref{fig1}(c). Considering this model, we have in mind two purposes. Firstly, the three mode scheme can be realized in practice as it is, i.e., by arranging a strong loss in the waveguide $c_0$. It is feasible, for example, in semiconductor planar waveguide structures \cite{balk}. Secondly, the three-mode model still allows for a comparison between the exact solution and the single-mode approximation for a moderately large numbers of photons (few tens). One could also use it to verify the linearization approximation, and check the optimal ratio of the interaction constants $g_b/g_a$ for minimizing the Mandel parameter with a fixed interaction time/length.

To verify the transition from the three modes model~\eqref{r3} to the single mode model~\eqref{r1},
the evolution of the mean photon number $\langle n_{{-}}(t) \rangle$ and the Mandel parameter $Q_{{-}}(t)$ of the antisymmetric collective mode $s_{{-}}$ was simulated using three modes model and quantum trajectories approach \cite{traj}. Obtained results were then compared with the same quantities computed using the single mode model with parameters taken from equations~\eqref{rates}. Figure~\ref{fig:ThreeModes_vs_SingleMode} illustrates how close these models are for the optimal ratio $g_b / g_a$~\eqref{optimum} and large enough losses in the third, lossy, mode. Solid lines in the figure correspond to the three modes model and dots correspond to the single mode model. The figure was obtained for  $g_b/g_a=\sqrt{2}-1$, $U = 0.012 g_b$, $\gamma_c = 6.04 g_b$, $\gamma_1 = 0$. 

Thus, the three modes model confirms the main evolution characteristic stemming from the single mode analysis, namely, rapid photon number squeezing for relatively small decrease in photon number in the initial stage of evolution.

\section{Multi-mode model}\label{sec:multi_mode}

Finally, we consider a complete multi-mode model depicted in Fig.~\ref{fig1}(a) for the realistic parameters described in Sec.~\ref{sec:singlemode}, and show that predictions made in that section still hold true for a multi-mode circuit of dissipatively coupled bosonic modes. We consider the multi-mode PhoG device with the two signal modes coupled to a long tail, which is described by Eq.~(\ref{rN}). 

We take the physical parameters from Sec.~\ref{sec:large_waveguides} for coupled waveguides in bulk glass, and the number of tail modes to be $N=28$. Since the Kerr nonlinearity parameter is small, $U \sim 10^{-8}$, in order to reach the regime of strongly sub-Poissonian light we have to take a bright input coherent state containing $\mathcal{O}\left(10^9\right)$ photons. To model such large numbers of photons and long tail lengths we recourse to the linearization method described in Sec.~\ref{sec:single_mode_linear} along the lines demonstrated in the Appendix~\ref{appendix:linear}.

\begin{figure}[htp]
\centering
\includegraphics[width=\linewidth]{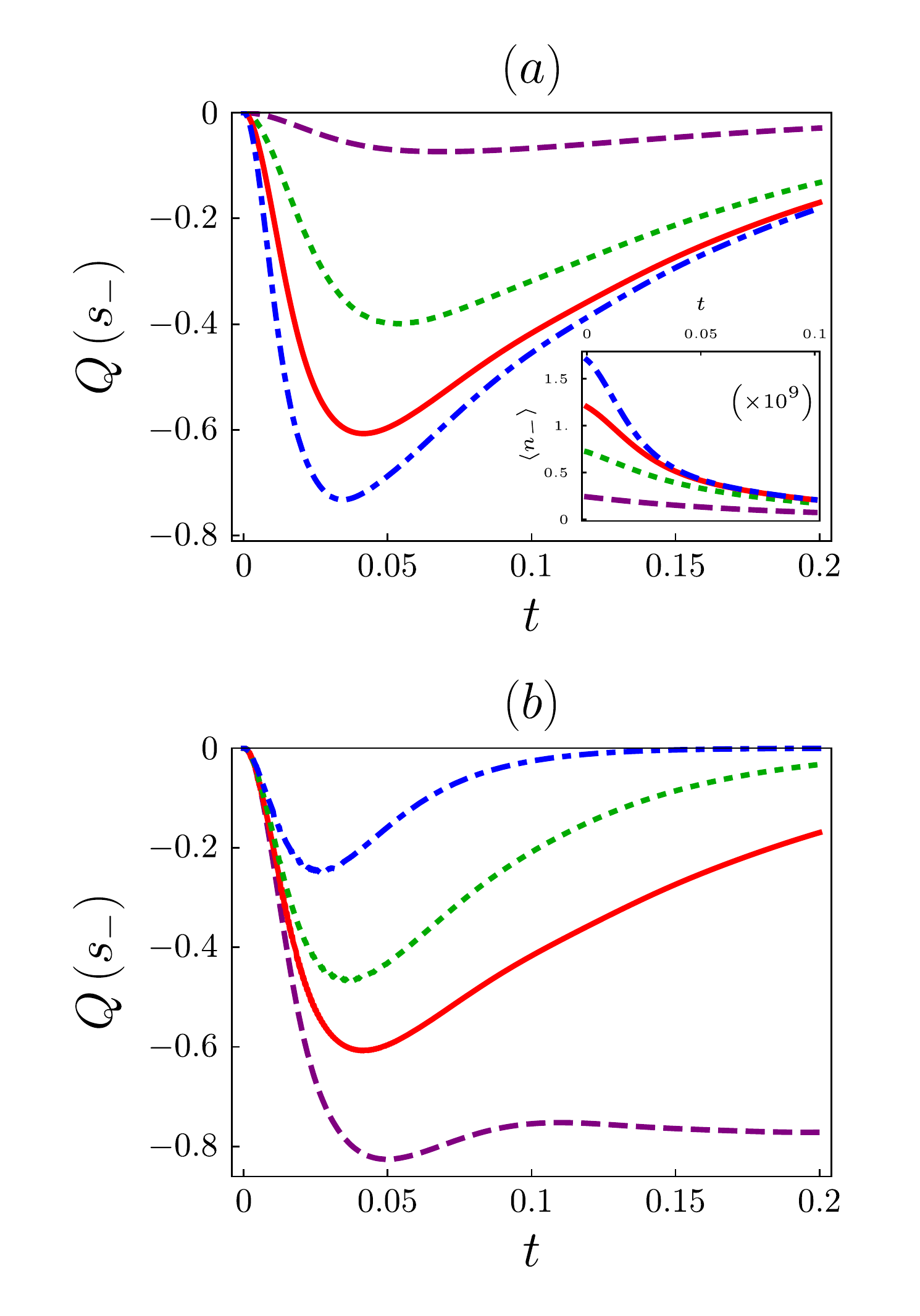}
\caption{\label{fig:multi} With the realistic parameters $U = 8.5\times10^{-8}$, $g_c = 60$m$^{-1}$, tail length $N=28$ and the optimal coupling ratio Eq.~\ref{optimum}, mode $s_-$ quickly evolves to a strongly sub-Poissonian state. A coherent state with average photon number $n_-\left(0\right)$ was initialized in mode $s_-$, and all other modes are initialized into the vacuum. (a) The dot-dashed, solid, dotted and dashed lines correspond to 
$\langle n_-\left(0\right)\rangle = 1.7\times 10^9, 1.2\times 10^9, 7.2\times10^8$ and $2.4\times 10^8$, respectively, and $\gamma_1 = 11.5$m$^{-1}$. \emph{Inset}: the evolution of photon number $\langle n_- \rangle$ displays the nonlinear decay behaviour perculiar to the NCL mechanism. (b) The Mandel parameter $Q$ remains strongly negative even in the presence of realistic linear loss $\gamma_1$. The dashed, solid, dotted and dot-dashed lines correspond to $\gamma_1 = 0, 11.5, 20$ and $40$m$^{-1}$, respectively, and $\langle n_- \left(0\right) \rangle = 1.2\times10^{9}$. }
\end{figure}

This linearization technique is applied to the full master equation, Eq.~(\ref{rN}). Since in the multi-mode model the Lindblad operators are only first-order in $a, b$ and $c_j$, only the third- and fourth- order linearization approximations are required. Therefore, we may reasonably expect the approximation to be more accurate in the multi-mode case than in the single-mode case, where linearization on sixth-order expectation values was required. We thus derive a closed system of equations for expectations up to second-order. An example of the evolution of both $Q$ and $\langle n_-\rangle$ is shown in Fig.~\ref{fig:multi}.

When a coherent state containing $1.2\times10^9$ photons is initialized in mode $s_-$ (solid red lines in Fig.~\ref{fig:multi}), the system quickly evolves to a strongly sub-Poissonian state, even in the presence of a realistic loss rate $\gamma_1$. A brighter initial state results in larger $|Q|$ over shorter timescales, and in greater robustness of the sub-Poissonian output to linear loss $\gamma_1$.

For no linear loss, $\gamma_1=0$ (dashed line in Fig.~\ref{fig:multi}, bottom), $Q \approx -0.8$ remains steady until a later time which is proportional to $N$. Thus, the tail effectively acts as a Markovian reservoir, which justifies its adiabatic elimination in Eq.~(\ref{r3}) \cite{natcom}.  Crucially, therefore, the long tail of modes combined with the self-Kerr interaction corresponds to the effects of nonlinear coherent loss into a Markovian reservoir (compare Fig.~\ref{fig:single_mode_exact} and Fig.~\ref{fig:multi}), and we note that the timescales over which this correspondence holds may be increased by increasing the tail length $N$. 

Since $s_-$ is a linear combination of modes $a$ and $b$, Eq.~(\ref{modes1}), the input coherent state with amplitude $\alpha$ into $s_-$ may be generated by initializing modes $a$ and $b$ with coherent states of appropriate amplitude. Alternatively, we may consider the initial condition $\langle a \rangle = \alpha$ and all other modes as vacuum. In this case, the behaviour of $s_-$ is qualitatively equivalent to Fig.~\ref{fig:multi}, but a slightly different optimal ratio $g_b/g_a$ should be considered.

Finally, we note that in the case of $\gamma_1 = 0$, the signal modes are highly sensitive to whether $N$ is odd or even, and for even $N$ the decay of mode $s_-$ is inhibited. Indeed, the decay rate $\gamma^\prime$ into an effective reservoir changes depending on $N$, and may be understood as an interference effect resulting from the unitary coupling between tail modes affecting whether the excitation returns to signal mode $a$ or mode $b$ from the tail. This effect of the reservoir, which quickly vanishes at even small levels of linear loss $\gamma_1$, will not manifest itself over the timescales of interest provided that $N$ is large. 

We have demonstrated that the multi-mode model -- which is the model most closely aligned to a physical implementation via waveguides inscribed in bulk glass, Fig.~\ref{fig1}(a) -- can generate a strongly sub-Poissonian output from a classical input state into mode $s_-$ over the short timescales of initial evolution. This corroborates results from Sec.~\ref{sec:singlemode} and the prediction of optimal coupling parameters $g_{a,b}$~\eqref{optimum} when mode $s_-$ is initially excited.

\begin{figure}[h!]
    \centering
            \includegraphics[width=1\linewidth]{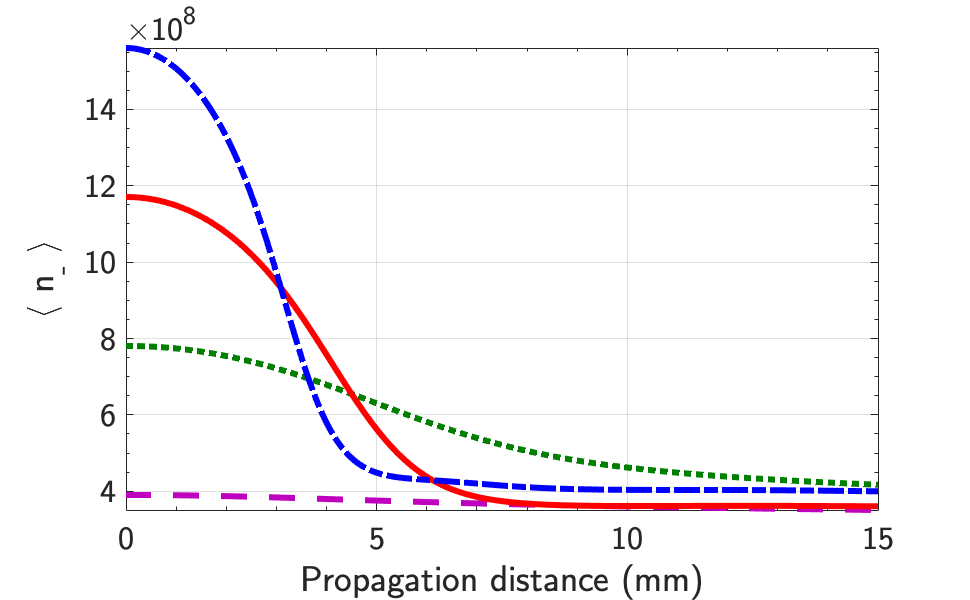}
    \caption{Evolution of photon number from the numerical solution of the $100$~fs pulse propagation in the waveguide structure with $5$ modes in the tail. We account explicitly for chromatic dispersion, Kerr effect and self-steepening, and solve the Nonlinear Schr{\"o}dinger equation via the split-step Fourier method for the system of coupled waveguides. The blue, red, green and purple colour lines correspond to $\langle n_{-}(0) \rangle =1.6\times10^{9},1.2\times10^9, 7.8\times10^8$ and $3.9\times10^8$. The evolution of the photon number characteristic for the NCL mechanism is observed (compare with Fig.~\ref{fig:multi}~(a)).}
    \label{fig:Optimal_coupling_energy_evolution}
\end{figure}

%%%%%% About pulse propagation
To assess the impact of specific effects of short-pulse propagation in the system of coupled waveguides, we dynamically evolved the spectral and temporal properties of the pulse and analysed their influence on the effectiveness of the NCL mechanism. In particular, we considered the combined effect of chromatic dispersion, self-phase modulation (Kerr effect), and self-steepening of the pulse via the Nonlinear Schr{\"o}dinger Equation (NLSE) for the system of waveguides in a $\chi^{\left(3\right)}$ nonlinear medium. The resulting system of coupled propagation equations for all waveguides in the PhoG structure was numerically solved using the split-step Fourier method \cite{Agrawal}. For a specific comparison we used the signature of the NCL mechanism, the nonlinear decay behaviour of photon number $\langle n_- \rangle$ in the anti-symmetric mode $s_-$ displayed in the inset of Fig.~\ref{fig:multi}~(a), which may additionally be interpreted as an intensity-dependent loss. In Fig.~\ref{fig:Optimal_coupling_energy_evolution} such decay can be clearly seen when evolving an initially Gaussian pulse according to the NLSE. Linear loss was neglected in this simulation to ensure that the decay was caused by the NCL mechanism, though we note that even when linear loss is included we observe strong agreement between the nonlinear decay behaviour of the full pulse propagation (NLSE) numerics and the quantum multi-mode model considered in Fig.~\ref{fig:multi}.

The pulse propagation was modelled for a $100$~fs initially Gaussian pulse in IG$2$ glass with a central wavelength of $1550$~nm (see Ref.\cite{ig2}, and also Sec.~\ref{sec:large_waveguides}). Initial pulse energies of $50$, $100$, $150$ and $200$~pJ were chosen, in order to compare with the inset of Fig.~\ref{fig:multi}~(a). The nonlinear parameter $\bar{\gamma}^{\rm NL}$ was set at $0.6~\textrm{W}^{-1}\textrm{m}^{-1}$, considering that the pulse energy was $200$~pJ. The resulting evolution is displayed in Fig.~\ref{fig:Optimal_coupling_energy_evolution} and we observe the characteristic nonlinear decay, which is a good evidence that NCL occurs for these realistic physical parameters even when the full pulse dynamics is considered.  Linear loss was neglected in this simulation to ensure that the decay was caused by NCL mechanism.

% In the numerical modelling, the number of modes in reservoir should be high enough to avoid the possibility for the field reaching the end of the tail to bounce back. Therefore high number of reservoir modes (modes $c_j$) is desirable, that however makes the problem computationally hard. As a compromise, five modes in the tail were considered in the simulation, giving a good degree of accuracy. The propagation was modelled for a $100$~fs Gaussian pulse in IG2 glass at $1550$~nm, as in Sec.~III, subsection C1. The nonlinear parameter $\bar{\gamma}^{NL}$ was set at $0.6~\textrm{W}^{-1}\textrm{m}^{-1}$, considering that the pulse energy was $200$~pJ.
%As a check-point, we used the signature of the NCL mechism, the nonlinear decay behavior of photon number $\langle n_- \rangle$ in the anti-symmetric mode $s_-$ displayed in the inset of Fig.~\ref{fig:multi}~(a). Linear loss was neglected in this simulation to ensure that the decay was caused by NCL mechanism. 
It was found that $g_a = 210\textrm{m}^{-1}$, $g_b = 360\textrm{m}^{-1}$ and $g_c = 430\textrm{m}^{-1}$ best reproduces the signature behaviour of NCL (different decaying rates for the different initial energies) as shown in Fig.~\ref{fig:Optimal_coupling_energy_evolution}. The optimal ratio between interaction constants is only weakly influenced by the  effects of the pulse propagation. The spectral broadening of the pulse is estimated to be about 10nm for 100 fs input pulse and 2-3 cm device at 1550 nm central wavelength. Provided that the device length is sufficient for the shortest wavelength of the pulse to undergo NCL decay (and for the example shown in Fig.~\ref{fig:Optimal_coupling_energy_evolution} it is sufficient),  the effects of such pulse spectral broadening do not affect the manifestations of the NCL mechanism.

% leads to different decaying rates for the different initial energies which reproduce best the signature behaviour of NCL, as shown in Fig.~\ref{fig:Optimal_coupling_energy_evolution}.

\section{Conclusions}

We have discussed a design of  a family of devices PhoG based of coupled single-mode waveguides in bulk nonlinear glass. We have shown that  an asymmetric structure can indeed function as a deterministic generator of bright sub-Poissonian states of light. The sub-Poissonian light generation with considerably  large photon number squeezing (up to the values of the Mandel parameter of about -0.8) can occur at the initial stages of the dynamics. In this regime conventional linear loss (which is quite high in nonlinear glass)  still can be overcome by sufficiently intensive coherent input, and the nonclassical states can be produced. A hierarchy of models has been derived, from the original multi-mode waveguide network down to the simplest single-mode model exhibiting both two-photon- and nonlinear-coherent- loss. We have developed an analytic approach to the single-mode model and discovered two dimensionless parameters allowing one to design the waveguide system with aim to reach a maximally nonclassical state for a given length of the structure. Feasibility of the suggested device has been analyzed for three different eligible practical structures: a system of waveguides in bulk glass, multi-core fiber, a set of coupled nano-wires. We have estimated the number of photons required to reach large nonclassicality with the femtosecond input pulses and demonstrated that this number remains well within the operational possibilities. 
Further, we have developed an approximation linearized on the quantum correction to the classical solution and demonstrated its closeness  with the exact solution of the single-mode problem. Both the three-mode approximation and the complete multi-mode model lead to very close results in the initial stage of the dynamics. 

In summary, suggested CDP circuits indeed allow for implemention of the engineered nonlinear loss to produce strongly non-classical states in realistic photonic structures. Our design of the PhoG with a system of waveguides laser-written in the bulk glass can serve as a practical recipe for realizing such structures. Being comparatively inexpensive and easy to produce, these generators can find applications for different detection, measurement and metrology tasks \cite{bio,imaging} and as quasi-single photon sources in various quantum technology applications, such as, e.~g., quantum key distribution \cite{wang1}. 

\section*{Acknowledgements}
The authors acknowledge support from the EU Flagship on Quantum Technologies, project PhoG (820365). D.M., A.S. and A. M. also acknowledge support from the National Academy of Sciences of Belarus program ``Convergence", and the BRRFI project F18U-006. We thank Benjamin Brecht for the help in developing the code for pulse propagation in waveguide network.
%\clearpage
\appendix 

\section{On Gaussianity of the photon-number distribution of the PhoG generated state}

Here we derive approximate analytical expressions for dynamics of mean photon number, Mandel parameter and higher-order moments of the photon number distribution for the single-mode model.  We start from the master equation (\ref{r1}) of the main text. 
For the photon number distribution, $p(n) = \langle n |\rho|n \rangle$, Eq. (\ref{r1}) yields
\begin{equation}
\label{p_equation}
\frac{d p(n)}{dz}= \sum_{i = 1}^3 \gamma_i \left[ f_i(n + \xi_i) p(n+\xi_i) - f_i(n) p(n) \right],
\end{equation}
where $\xi_i$ is the number of the photons lost per each dissipation event ($\xi_1 = \xi_3 = 1$, $\xi_2 =2$); $f_1(n) = n$, $f_2(n) = n (n-1)$, and $f_3(n) = n(n-1)^2$.

For a function $g(n)$, one can define its average as
\begin{equation}
\left\langle g(n) \right\rangle = \sum_{n=0}^\infty g(n) p(n).
\end{equation}
Eq. (\ref{p_equation}) implies that the average satisfies the following equation:
\begin{equation}
\label{average_derivative}
\frac{d}{dz} \left\langle g(n) \right\rangle =  \sum_{i = 1}^3 \gamma_i \left\langle f_i(n) \left[ g(n - \xi_i) - g(n) \right] \right\rangle.
\end{equation}

For example, the mean photon number $\mu = \langle n \rangle$ corresponds to $g(n) = n$ and satisfies
\begin{equation}
\label{mu_derivative}
\begin{gathered}
\frac{d \mu}{dz} = - \sum_{i = 1}^3 \gamma_i \xi_i \langle f_i(n) \rangle \\ {} = - \gamma_1 \mu - 2 \gamma_2 \left(\langle n^2 \rangle - \mu \right) - \gamma_3 \left(\langle n^3 \rangle - 2 \langle n^2\rangle + \mu \right).
\end{gathered}
\end{equation}

For further calculation of the photon number distribution moments, it is also useful to calculate the derivative for a function $g(\delta n)$ of the photon number deviation $\delta n = n - \mu$. Here, one needs to take into account that $\mu$ also varies with $z$:
\begin{equation}
\label{average_dn_derivative}
\begin{gathered}
\frac{d}{dz} \left\langle g(\delta n) \right\rangle =  \sum_{i = 1}^3 \gamma_i \left\langle f_i(n) \left[ g(\delta n - \xi_i) - g(\delta n) \right] \right\rangle \\ {} -\left\langle g'(\delta n) \right\rangle \frac{d \mu}{dz} \\{} =  \sum_{i = 1}^3 \gamma_i \{ \left\langle f_i(n) \left[ g(\delta n - \xi_i) - g(\delta n) \right] \right\rangle \\{} + \xi_i \left\langle g'(\delta n) \right\rangle \left\langle f_i(n) \right\rangle \}.
\end{gathered}
\end{equation}
In further calculations, $n$ in Eq. (\ref{average_dn_derivative}) can also be expressed in terms of $\delta n$: $n = \mu + \delta n$.

Let us assume that the initial state is a coherent one with the mean photon number $\mu(z=0) = n_0$:
\begin{equation}
p(n, z=0) = \frac{n_0^n}{n!} e^{-n_0}.
\end{equation}
Central moments of the input state are equal to
\begin{equation}
\label{moments_initial}
\langle \delta n^2 \rangle = \mu, \quad \langle \delta n^3 \rangle = \mu, \quad \langle \delta n^4 \rangle = 3 \mu^2 + \mu. 
\end{equation}
Taking into account that Poisson distribution does not have heavy tails and its central moments are mainly determined by the central part of the distribution, one may expect that, at least for the beginning of the state evolution, $p(n)$ will have non-zero values for $|\delta n| \lesssim const \sqrt \mu$, and, therefore, 
\begin{equation}
\label{higher_moments_initial}
\langle \delta n^k \rangle = O(\mu^{k/2}).
\end{equation}

Let us consider such $z$ that the conditions (\ref{higher_moments_initial}), together with
\begin{equation}
\label{zeta2_order}
\zeta_2 = \frac{\langle \delta n^2 \rangle}{\mu} = O(1),
\end{equation}
are still satisfied. Substituting definition of $\zeta_2$ into Eq. (\ref{average_dn_derivative}) and using Eq. (\ref{higher_moments_initial}) for $k \ge 3$, one can derive the following equation:
\begin{equation}
\label{zeta2_equation}
\begin{gathered}
\frac{d\zeta_2}{dz}=\gamma_1 \left(1 - \zeta_2\right) \\{} + 2 \gamma_2  \left\{ \left(2 -3 \zeta_2 \right) \mu + O(\sqrt \mu) \right\} \\ {} + \gamma_3 \left\{ \left(1 - 5 \zeta_2\right) \mu^2 +O(\mu^{3/2})\right\}.
\end{gathered}
\end{equation}

According to the derived equation, the quantity $\zeta_2$ tends to reach the value
\begin{equation}
\label{zeta2_limit}
\zeta_2^{(f)} \approx \frac{\gamma_1 + 4 \gamma_2 \mu + \gamma_3 \mu^2}{\gamma_1 + 6 \gamma_2 \mu + 5 \gamma_3 \mu^2} = O(1).
\end{equation}
Taking into account that $\zeta_2(z=0) = 1$, one can expect that the condition (\ref{higher_moments_initial}) will be satisfied during most of the system evolution, until the mean photon number $\mu$ becomes small.

Assuming that a single type of dissipation is prevailing, one can derive the values
\begin{equation}
\zeta_2^{(f)} = 1, \; 2/3,\; 1/5
\end{equation}
for linear, two-photon, and nonlinear coherent loss respectively. The corresponding values of Mandel parameter $Q = \zeta_2 - 1$ are
\begin{equation}
Q = 0, \; -1/3,\; -4/5.
\end{equation}

Now let us assume that for the considered $z$ the conditions
\begin{equation}
\label{zeta3_order}
\zeta_3 = \frac{\langle \delta n^3 \rangle}{\mu} = O(1)
\end{equation}
and
\begin{equation}
\label{zeta4_order}
\zeta_4 = \frac{\langle \delta n^4 \rangle - 3 \zeta_2 \mu}{\mu^{3/2}} = O(1)
\end{equation}
are still satisfied (for $z=0$ one has $\zeta_3 = 1$ and $\zeta_4 = 1/\sqrt \mu$) together with Eq. (\ref{higher_moments_initial}) for $k \ge 5$.

Therefore, one can derive the following equation for $\zeta_3$:
\begin{equation}
\label{zeta3_equation}
\begin{gathered}
\frac{d\zeta_3}{dz}=\gamma_1 \left(- 1 +3 \zeta_2 - 2\zeta_3\right) \\{} + 2 \gamma_2  \left\{ \left(-4 + 12 \zeta_2 - 6 \zeta_2^2 - 5 \zeta_3 \right) \mu + O(\sqrt \mu) \right\} \\ {} + \gamma_3 \left\{ \left(-1 +9\zeta_2 - 18 \zeta_2^2 - 8 \zeta_3\right) \mu^2 +O(\mu^{3/2})\right\}.
\end{gathered}
\end{equation}

The quantity $\zeta_3$ tends to the value
\begin{equation}
\label{zeta3_limit}
\begin{gathered}
\zeta_3^{(f)} \approx \frac{1}{2 \gamma_1 + 10 \gamma_2 \mu + 8 \gamma_3 \mu^2} [\gamma_1 (3 \zeta_2 - 1) \\{}+ 4 \gamma_2 \mu (-2 +6 \zeta_2 - 3 \zeta_2^2) \\{}+ \gamma_3 \mu^2 (-1 + 9 \zeta_2 - 18 \zeta_2^2)] = O(1).
\end{gathered}
\end{equation}

In a similar way one can show that $\zeta_4$ satisfies
\begin{equation}
\label{zeta4_equation}
\begin{gathered}
\frac{d\zeta_4}{dz}=\gamma_1 \left\{- 4 \zeta_4 \sqrt\mu + O(1)\right\} \\{} + 2 \gamma_2  \left\{ -8 \left(2 \zeta_4 + \frac{\langle \delta n^5 \rangle}{\mu^{5/2}} \right) \mu^{3/2} + O(\mu) \right\} \\ {} + \gamma_3 \left\{-12  \left(\zeta_4 + + \frac{\langle \delta n^5 \rangle}{\mu^{5/2}} \right) \mu^{5/2} +O(\mu^2)\right\}.
\end{gathered}
\end{equation}

The derived equation has two important implications. First, it shows that the condition $\zeta_4 = O(1)$ remains valid while $\langle \delta n^5 \rangle / \mu^{5/2} = O(1)$, which in its turn is a consequence of Eqs. (\ref{zeta2_equation}) and (\ref{zeta2_limit}). Second, in contrast to lower-order moments, the equation for $\zeta_4$ is not disentangled from higher moments and cannot be solved independently of them. Still, the condition $\zeta_4 = O(1)$ is sufficient for validity of the previously derived Eqs. (\ref{zeta3_equation}) and (\ref{zeta2_limit}).

For a single type of dissipation, one can derive the values
\begin{equation}
\zeta_3^{(f)} = 1, \; 2/30,\; 1/100
\end{equation}
for linear, two-photon, and nonlinear coherent loss respectively, which correspond to the following values of skewness $\zeta_3 / (\sqrt \mu \zeta_2^{3/2})$:
\begin{equation}
\mbox{Skewness} = \frac{1}{\sqrt \mu},\; \frac{0.12}{\sqrt \mu},\; \frac{0.11}{\sqrt \mu}.
\end{equation}
Similarly, for excess kurtosis one can derive the estimate
\begin{equation}
\mbox{Excess kurtosis} = \frac{\langle \delta n^4 \rangle}{\mu^2 \zeta_2^2} - 3 = \frac{\zeta_4}{\sqrt \mu \zeta_2^2} = O\left(\frac{1}{\sqrt \mu}\right).
\end{equation}

The results show that the photon number distribution, despite becoming strongly sub-Poissonian, remain Gaussian with high accuracy until the mean photon number $\mu$ becomes small.

\section{Derivation of linearization approximations}\label{appendix:linear}
In this section we will derive the linearization approximations which are used in Sec.~\ref{sec:single_mode_linear} and Sec.~\ref{sec:multi_mode}. These approximations become increasingly necessary for modelling the system when large photon numbers or large numbers of modes must be considered.

Consider arbitrary quantum operators $A, B, C$. We will explicitly show the derivation for linearization of $\langle A B C \rangle$, but the approach may be readily generalized to higher-order products. We will derive replacements which will allow us to approximate expectations of products of three or more operators, using only first- and second- order terms e.g. $\langle A \rangle$, $\langle A B \rangle$.

We may expand each operator $A, B, C$ into a classical ``mean-field" term and a quantum fluctuation term: $A = \langle A \rangle + \delta A$. We take $\langle \delta A \rangle = 0$ which ensures that the mean-field term $\langle A \rangle$ is meaningful and consistent. Then substituting these expansions into $\langle A B C \rangle$, 

\begin{align}\label{eqn:Aexpansion}
\langle A B C \rangle = \langle A \rangle \langle B \rangle \langle C \rangle + \langle A \rangle \langle \delta B \delta C \rangle + \langle B \rangle \langle \delta A \delta C \rangle& \notag \\+ \langle C \rangle \langle \delta A \delta B \rangle + \langle \delta A \delta B \delta C \rangle&.
\end{align}

A key tool we require is the cumulant expansion for generic operators $\left\{O_1,\dots, O_n\right\}$

\begin{equation}
\mathcal{C}\left(O_1,\dots,O_n\right) = \sum_{\mathcal{P} \in \mathbb{P}}\left(|\mathcal{P}| - 1\right)!\left(-1\right)^{|\mathcal{P}|-1}\prod_{p \in \mathcal{P}}\langle \prod_{i \in p} O_i\rangle
\end{equation}

where $\mathbb{P}$ denotes all disjoint partitions of the set of operators, $|\mathcal{P}|$ denotes the number of blocks in partition $\mathcal{P}$, and $p$ iterates over each block in the partition. For example

\begin{align*}
\mathcal{C}\left(X, Y, Z\right) = \langle X Y Z \rangle + 2 \langle X \rangle \langle Y \rangle \langle Z \rangle - \langle X \rangle \langle Y Z \rangle& \notag \\
 - \langle Y \rangle \langle X Z \rangle - \langle Z \rangle \langle X Y \rangle&.
\end{align*}

We perform our linearization assumption on Eq.~\ref{eqn:Aexpansion} by setting $\mathcal{C}\left( \delta A, \delta B, \delta C\right) = 0$, which implies therefore that $\langle \delta A \delta B \delta C \rangle = 0$ since $\langle \delta A \rangle = \langle \delta B \rangle = \langle \delta C \rangle = 0$. Finally, using $\delta A = A - \langle A \rangle$ we arrive at our final expression,

\begin{align}
\langle A B C \rangle \approx &\langle A \rangle \langle B C \rangle + \langle B \rangle \langle A C \rangle + \langle C \rangle \langle A B \rangle \notag \\
&- 2 \langle A \rangle \langle B \rangle \langle C \rangle 
\end{align}

Higher-order expectations of products of operators may be calculated in the same way, with the only requirements assumed about fluctuations being the zero-mean condition $\langle \delta A \rangle = \dots = \langle \delta Z \rangle = 0$ and the linearization approximation $\mathcal{C}\left(\delta A, \dots, \delta Z\right) = 0$.

%\clearpage
\section{Linearized single-mode model}\label{appendix:single_mode_linear}
By applying the linearization approximations derived in Appendix~\ref{appendix:linear} to the system of coupled ODEs \eqref{eqn:expectations} we arrive at the following closed system of ODEs:

\begin{align}\label{eqn:expectations_linear_first}
%
%a
%
\partial_t\langle s_- \rangle &= c_1 \langle s_- \rangle + c_2 \left(\langle s_-^\dagger \rangle \langle s_-^2 \rangle + 2 \langle s_- \rangle \langle n_-\rangle - 2 \langle s_-^\dagger \rangle \langle s_- \rangle^2\right) \notag \\
&-\frac{\gamma_3}{2}\left(
6 \langle s_-^\dagger \rangle \langle n_- \rangle \langle s_-^2\rangle + 3 \langle s_- \rangle \langle s_-^{\dagger 2} \rangle \langle s_-^2 \rangle \right. \notag \\
&+ 6 \langle s_- \rangle \langle n_- \rangle^2 - 2 \langle s_-^{\dagger 2} \rangle \langle s_- \rangle^3 - 12 \langle n_- \rangle \langle s_-^\dagger \rangle \langle s_- \rangle^2 \notag \\
&\left. - 6 \langle s_-^2 \rangle \langle s_-^\dagger \rangle^2 \langle s_- \rangle + 6 \langle s_-^\dagger \rangle^2 \langle s_- \rangle^3\right)
\end{align}
\begin{align}
%
% ad
%
\partial_t\langle s_-^\dagger \rangle &= c_1^* \langle s_-^\dagger \rangle + c_2^* \left(\langle s_- \rangle \langle s_-^{\dagger 2} \rangle + 2 \langle s_-^\dagger \rangle \langle n_- \rangle - 2 \langle s_-^\dagger \rangle^2\langle s_- \rangle \right) \notag \\
&-\frac{\gamma_3}{2}\left(6 \langle s_- \rangle \langle n_- \rangle \langle s_-^{\dagger 2}\rangle + 3 \langle s_-^\dagger \rangle \langle s_-^2 \rangle \langle s_-^{\dagger 2}\rangle \notag \right. \notag \\
&+ 6 \langle s_-^\dagger \rangle \langle n_-\rangle^2 - 2 \langle s_-^2 \rangle \langle n_-^3\rangle - 12 \langle n_- \rangle \langle s_- \rangle \langle s_- \rangle^2 \notag \\
&\left. - 6 \langle s_-^{\dagger 2} \rangle \langle s_- \rangle^2 \langle s_-^\dagger \rangle + 6 \langle s_-^{\dagger 3} \rangle \langle s_- \rangle^2 \right)
\end{align}
\begin{align}
%
% aa
%
\partial_t\langle s_-^2 \rangle &= c_3 \langle s_- s_- \rangle + c_4 \left( 3 \langle n_- \rangle \langle s_-^2 \rangle  - 2 \langle s_-^\dagger \rangle \langle s_- \rangle^3 \right) \notag \\
&- \gamma_3 \left( 3 \langle s_-^{\dagger 2}\rangle \langle s_-^2\rangle^2 + 12 \langle n_- \rangle^2 \langle s_-^2\rangle - 2 \langle s_-^{\dagger 2}\rangle \langle s_- \rangle^4 \right. \notag \\
& - 12 \langle s_-^2 \rangle \langle s_-^{\dagger}\rangle^2 \langle s_- \rangle^2 - 16 \langle n_- \rangle \langle s_-^\dagger \rangle \langle s_- \rangle^3 \notag \\
&\left. + 16 \langle s_-^\dagger \rangle^2 \langle s_- \rangle^4 \right)
\end{align}
\begin{align}
%
% adad
%
\partial_t\langle s_-^{\dagger^2} \rangle &= c_3^* \langle s_-^{\dagger 2}\rangle + c_4^* \left(3 \langle s_-^{\dagger 2} \rangle \langle n_- \rangle - 2 \langle s_-^\dagger \rangle^3 \langle s_- \rangle \right) \notag \\
&- \gamma_3 \left(3 \langle s_-^{\dagger 2} \rangle^2 \langle s_-^2 \rangle + 12 \langle s_-^{\dagger 2} \rangle \langle n_- \rangle^2 - 2\langle s_-^2 \rangle \langle s_-^\dagger \rangle^4 \right. \notag \\
&\left. - 21 \langle s_-^{\dagger 2}\rangle \langle s_-^\dagger\rangle^2 \langle s_- \rangle^2 - 16 \langle n_- \rangle \langle s_-^\dagger \rangle^3 \langle s_-\rangle \right. \notag \\
& \left.+ 16 \langle s_-^\dagger\rangle^4 \langle s_- \rangle^2 \right)
\end{align}
\begin{align}\label{eqn:expectations_linear_last}
%
% ada
%
\partial_t \langle n_- \rangle &= - \gamma_1 \langle n_- \rangle + c_5 \left( \langle s_-^{\dagger 2} \rangle \langle s_-^2 \rangle + 2 \langle n_- \rangle^2 - 2 \langle s_-^\dagger \rangle^2 \langle s_- \rangle^2 \right) \notag \\
&- \gamma_3 \left( 9 \langle s_-^{\dagger 2} \rangle \langle n_-  \rangle \langle s_-^2\rangle + 6 \langle n_-\rangle^3  - 6 \langle s_-^{\dagger 2}\rangle \langle s_-^\dagger \rangle \langle s_- \rangle^3 \right. \notag \\
& - 18 \langle n_- \rangle \langle s_-^\dagger \rangle^2 \langle s_- \rangle^2 - 6 \langle s_-^2 \rangle \langle s_-^\dagger \rangle^3 \langle s_- \rangle \notag \\
& \left. + 16 \langle s_-^\dagger \rangle^3 \langle s_-\rangle^3 \right)
\end{align}
with $n_- = s_-^\dagger s_-$.

This system is solved numerically for $\langle s_- \rangle$, $\langle s_-^\dagger \rangle$, $\langle s_-^2\rangle$, $\langle s_-^{\dagger 2}\rangle$, $\langle n_-\rangle$ and the results are shown as dashed lines in Fig.~\ref{fig:Qlinvcompare}.

\end{document}